\documentclass[10pt,journal,compsoc]{IEEEtran}
\usepackage{amssymb}
\usepackage{booktabs} 
\usepackage{balance}
\usepackage{mathtools}
\usepackage{subfig}
\usepackage{graphicx}
\usepackage{amsmath}
\usepackage{xcolor}
\usepackage{tikz}
\usepackage{enumitem,kantlipsum}
\usepackage{epstopdf}

\ifCLASSOPTIONcompsoc
  \usepackage[nocompress]{cite}
\else
  \usepackage{cite}
\fi
\ifCLASSINFOpdf
\else
\fi
\begin{document}
\bstctlcite{IEEEexample:BSTcontrol}
%
\title{Local Differentially Private Fuzzy Counting in Stream Data using Probabilistic Data Structures}
%
%
%
%

\author{Dinusha~Vatsalan,~
        Raghav~Bhaskar, 
        and~Mohamed~Ali~Kaafar 
\IEEEcompsocitemizethanks{\IEEEcompsocthanksitem Dinusha Vatsalan is 
with the Faculty of Science and Engineering, Macquarie University, Sydney, Australia, \protect\\
Email: dinusha.vatsalan@mq.edu.au 
\IEEEcompsocthanksitem Raghav Bhaskar is with 
AppsPicket, New Delhi, India, \protect\\
E-mail: raghav@appspicket.com
\IEEEcompsocthanksitem Mohamed Ali Kaafar is with the Faculty of Science and Engineering, Macquarie University, 
Sydney,  Australia, \protect\\
E-mail:dali.kaafar@mq.edu.au
}
\thanks{Manuscript received May 03, 2022; revised August 09, 2022.}}

%
%

\newcommand{\theHalgorithm}{\arabic{algorithm}}
\newtheorem{definition}{Definition}[section]
\newtheorem{theorem}{Theorem}[section]
\newtheorem{prop}[theorem]{Proposition}
\newtheorem{proof}[theorem]{Proof}

\markboth{Local Differentially Private Fuzzy Counting in Stream Data}%
{Shell \MakeLowercase{\textit{et al.}}: Local Differentially Private Fuzzy Counting in Stream Data}
%



\IEEEtitleabstractindextext{%
\begin{abstract}
Privacy-preserving estimation of counts of items in streaming data finds applications in several real-world scenarios including 
word auto-correction and traffic management applications.
Recent works of RAPPOR~\cite{Erl14} and Apple’s count-mean sketch (CMS) algorithm~\cite{Dp17} propose privacy preserving mechanisms for count estimation in large volumes of data using probabilistic data structures like counting Bloom filter and CMS. 
However, these existing methods fall short in providing a sound solution for real-time streaming data applications.
Since the size of the data structure in these methods is not adaptive to the volume of the streaming data, the utility (accuracy of the count estimate) can suffer over time due to increased false positive rates. Further, the lookup operation needs to be highly efficient to answer count estimate queries in real-time. More importantly, the local Differential privacy mechanisms used in these approaches to provide privacy guarantees come at a large cost to utility (impacting the accuracy of count estimation).

In this work, we propose a novel (local) Differentially private mechanism that provides high utility for the streaming data count estimation problem with similar or even lower privacy budgets while providing: a) fuzzy counting to report counts of related or similar items (for instance to account for typing errors and data variations), and b) improved querying efficiency to reduce the response time for real-time querying of counts. 
Our algorithm uses a combination of two probabilistic data structures Cuckoo filter and Bloom filter. 
We provide formal proofs for privacy and utility guarantees and present extensive experimental evaluation of our algorithm using real and synthetic English words datasets for both the exact and fuzzy counting scenarios. Our privacy preserving mechanism substantially outperforms the prior work in terms of lower querying time, significantly higher utility (accuracy of count estimation) under similar or lower privacy guarantees, at the cost of communication overhead.
\end{abstract}

}

\maketitle

\IEEEdisplaynontitleabstractindextext

%
\IEEEpeerreviewmaketitle

\IEEEraisesectionheading{\section{Introduction}\label{sec:Introduction}}

\IEEEPARstart{T}{he} growing demand for real-time and faster analytics of continuously generated big volumes of data brings tremendous interest in streaming data technologies. However, privacy concerns in sharing or revealing personal information require privacy preserving techniques to be designed for such technologies. Privacy preserving counting of frequency of items in the data is useful in several real-time streaming data applications. 

An example application is word auto-correction that requires learning the frequency counts of word entries by many different users such that when a user enters a common/frequent word (e.g. LOL, according to a frequency threshold), it will not be auto-corrected~\cite{Dp17}. Another example is Web data obfuscation application for privacy preserving Web browsing, which requires counting the uniqueness of data entry or click path entered by all users in order to calculate privacy risks and obfuscate accordingly in real-time~\cite{Mas18}. 
In such applications, data entries need to be continuously monitored from many users~\footnote{note that the frequency needs to be calculated across multiple users, i.e. how many users have entered a certain data, not the frequency of a single user} and the auto-correction or auto-obfuscation function needs to query for the frequency count information from many users in real-time and on-the-fly when a user actively enters/types a data/word in an application.  
However, users' individual data entries may identify their private and sensitive information, for example, personal interests, occupation, health, or location, thereby necessitating the use of privacy preserving techniques in the counting task.

There have been few methods proposed for the count-frequency problem in streaming data including the two state-of-the-art methods with provable privacy guarantees: RAPPOR (Randomized Aggregatable Privacy-Preserving Ordinal Response) proposed by Google for collecting statistics from end-user client software~\cite{Erl14} and the count-mean sketch-based approach introduced by Apple to discover frequency of words or emojis used by users~\cite{Dp17}\footnote{From now onwards, we may refer to the count-mean sketch-based approach introduced by Apple in \cite{Dp17} as Apple's algorithm}. 
However, these methods are developed for offline processing of count querying functions used for analytics purposes and not tailored for real-time or online query processing.
We aim to improve real-time processing for count querying functions where the count queries need to be answered in real-time and online. For example, in the auto-spelling correction and privacy-aware Web data obfuscation applications (described above), the decision to auto-correct words and obfuscate data needs to be retrieved by the corresponding applications in near real-time when a user types a word in a mobile App or enters data in the Web (e.g. a search query)~\cite{Mas18}.

Given sufficient time and resources, calculating frequency of items is a simple task, i.e. just keeping a count of observations for each item in the data to obtain that item's frequency. However, in the context of high-scale, low-latency, and online data processing, counts of items need to be calculated instantly as the query comes in. 
The naive approach of randomly sampling the observations for estimating the counts with the assumption that the sample generally reflects the properties of the whole is not effective, as ensuring true randomness is a difficult task. That is where probabilistic data structures come in to estimate the approximate counts of items in an efficient way~\cite{Fan14,Mit05,Sch15}. 
Probabilistic data structures generally trade space and computational efficiency for accuracy (false positive rate) of data processing.

Google's RAPPOR and Apple's algorithm use the probabilistic data structures, counting Bloom filter and count-mean sketch, respectively, for efficient privacy preserving counting~\cite{Erl14,Dp17}. For real-time count querying applications, the lookup operational time is most crucial in order to answer the queries in near real-time.
The lookup operations in these probabilistic data structures are dependent on the number of hash functions used and the size of the probabilistic data structures. Reducing the number of hash functions and the size of these probabilistic data structures can improve the lookup operation in terms of query response time and space, however, it comes at a cost of utility loss.  
Further, continuous insertion of data into these probabilistic data structures leads to increasing probability of collisions of different items and therefore impacts the false positive rate that eventually reaches to $1.0$. In fact, our experimental results on large English words datasets show that the false positive rate with the RAPPOR and CMS methods reaches to $1.0$ when approximately $100,000$ records are inserted (as presented in Section~\ref{sec:experiments}). 

Moreover, in the example applications (described above), the same data entered/typed by different users might have different variations or forms (e.g. lemma) or errors (e.g. typos), and therefore exact counting does not provide good utility of count estimation in these applications. Existing works in privacy preserving counting allow only exact matching of items when querying for frequency counts of items~\cite{Dp17,Erl14}. 
Both~\cite{Erl14} and~\cite{Dp17} achieve local Differential privacy guarantees for the data items contributed by the clients by using variants of the Randomized Response~\cite{Hol17} techniques. A key challenge of using Differential privacy technologies is achieving a good balance between privacy and utility guarantees. 

In this paper, we propose a novel Differentially private counting algorithm for real-time streaming applications that addresses all the above-described limitations or shortcomings of the existing methods.

\noindent
\textbf{Contributions:}
\begin{enumerate}
    \item We propose a Differentially private Cuckoo filter for efficient counting. The lookup operation with cuckoo filter does not depend on the number of hash functions nor the size of the filter, and thus is more efficient than other probabilistic data structures in terms of lookup time and space~\cite{Fan14}.

\item We overcome the issue of the impact of velocity of data on the utility (false positive probability) by extending the  cuckoo filter to be adaptive to large volume of data that keeps the false positive rate bounded regardless of the number of items inserted. 

\item Our algorithm allows efficient and effective fuzzy counting. We combine Bloom filters~\cite{Bloom70} with cuckoo filters to allow fuzzy matching for counting of values that are similar to each other or in a similar range. 

\item In order to reduce the loss in utility due to noise addition, i.e., to achieve better trade-off between privacy and utility, we propose a novel local Differential privacy mechanism for our method that instead of perturbing the Bloom filter (encoded item) itself, adds noise to generate "artificial" Bloom filters as noise such that an adversary is unable to distinguish a real Bloom filter from "artificial" ones.
In our method, noise is introduced to clients' input via two steps: choosing the bucket positions of the cuckoo filter and then choosing the corresponding Bloom filters that need to be sent for each of those buckets. At most one of these Bloom filters comes from the real item and all others correspond to “artificial” items.  
We provide formal proofs for the privacy and utility guarantees of our method.

\item Using real and synthetic English word frequency datasets, we conduct an experimental study of our method and compare it to two state-of-the-art methods, Google's RAPPOR~\cite{Erl14} and Apple's algorithm~\cite{Dp17}. Our experimental results show that our method outperforms RAPPOR and Apple's algorithm by a large margin in terms of significantly higher accuracy of count estimation with both exact and fuzzy counting cases (around 60\% higher) and lower querying time (more than one order magnitude lower cumulative time) with similar privacy guarantees and insertion/update time. Since answering count queries instantly and more accurately in online count querying applications (e.g. word auto-correction or privacy-aware data obfuscation applications) is critical, our method highly outperforms the state-of-the-art methods for such applications.
\end{enumerate}

\noindent
\textbf{Outline:}
We provide preliminaries in the following section and describe our methodology in Section~\ref{sec:methodology}. In Section~\ref{sec:experiments} we present the results of our experimental study and in Section~\ref{sec:rw} we review the literature of privacy preserving counting and Cuckoo filter techniques. Finally we conclude and provide directions to future research in Section~\ref{sec:conclusion}.

\begin{figure*}[th]
    \centering
    \includegraphics[width= 0.33\linewidth, keepaspectratio]{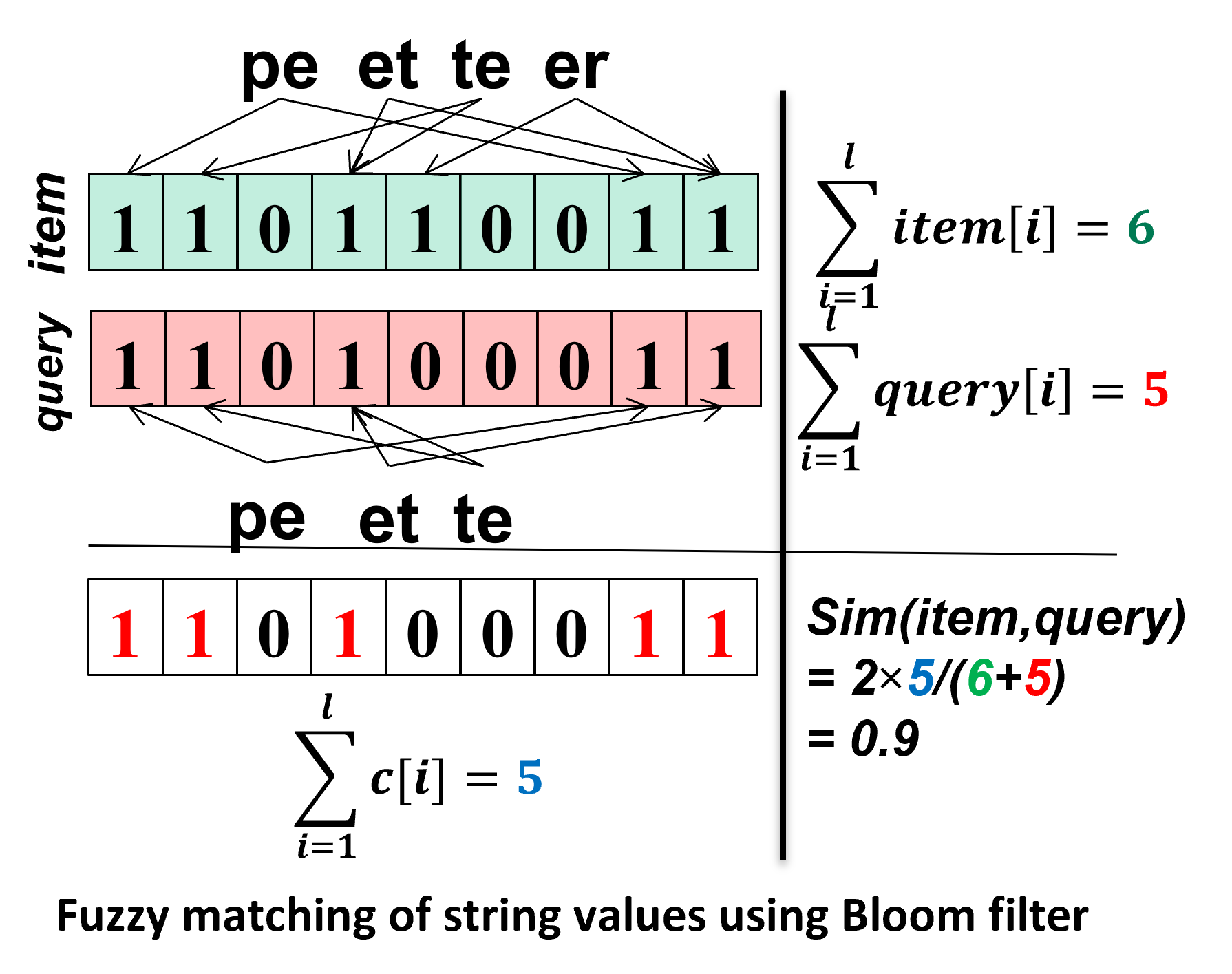}
    \includegraphics[width= 0.33\linewidth, keepaspectratio]{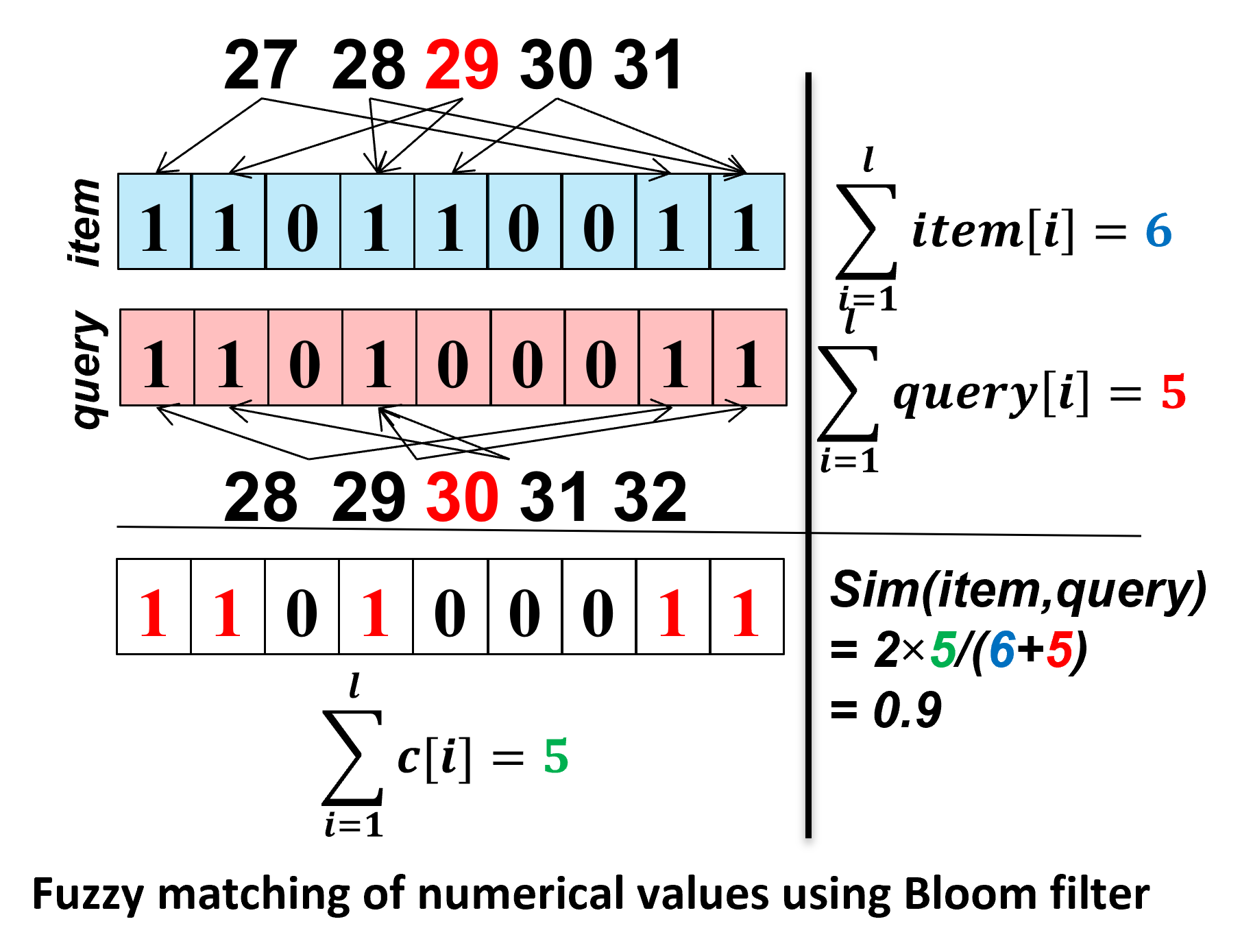}
    \includegraphics[width= 0.33\linewidth, keepaspectratio]{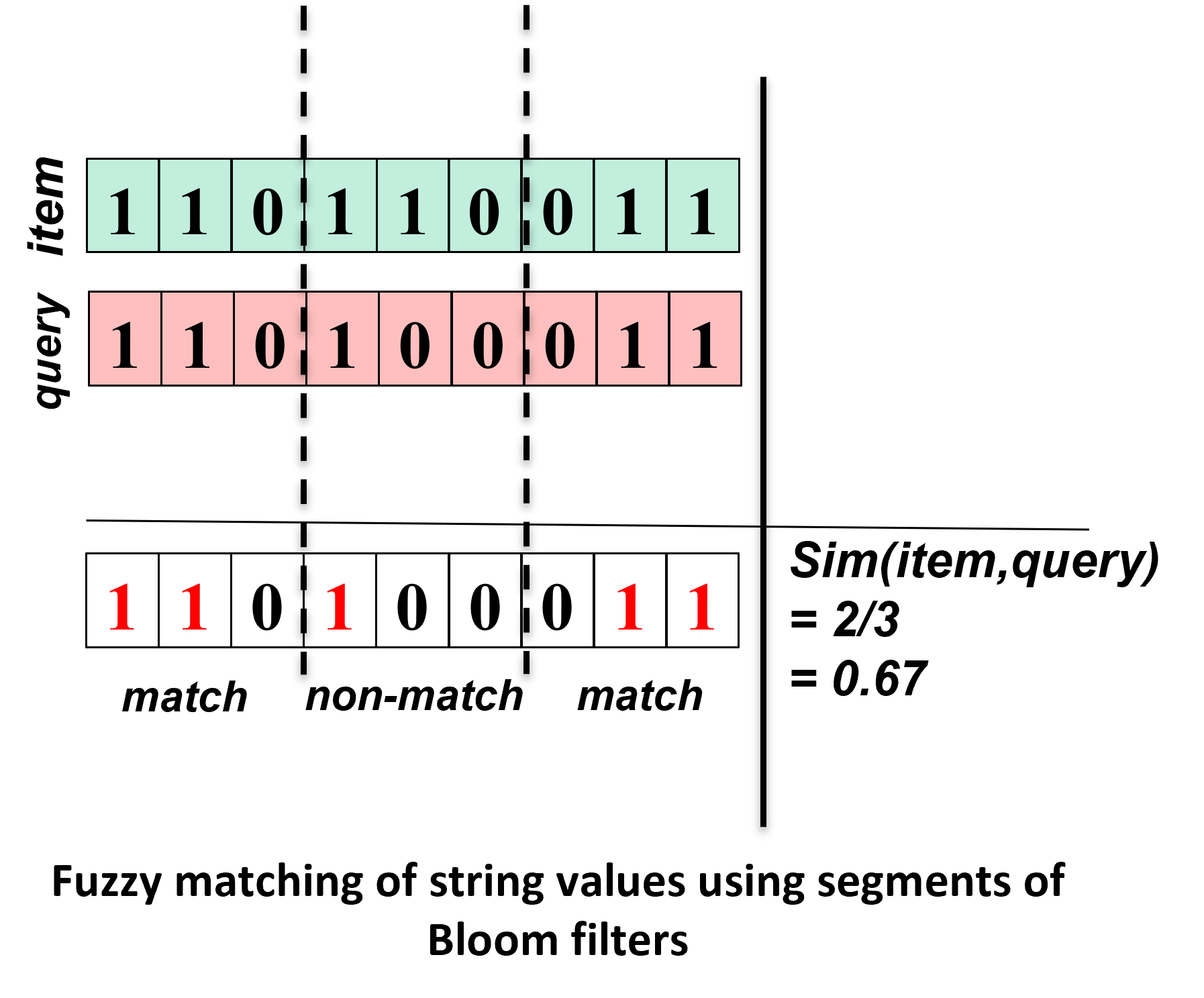}
    \caption{Fuzzy counting of string values (left) and numerical values (middle) using Bloom filter encoding~\cite{Sch15,Vat16}, and fuzzy matching of Bloom filter segments (right) used in our proposed fuzzy counting method described in detail in Section~\ref{subsec:fuzzy_matching}.}
    \label{fig:BFs}
\end{figure*}

\section{Preliminaries}
\label{sec:preliminaries}

In this section, we describe some preliminaries of probabilistic data structures in use, fuzzy counting, and the two state-of-the-art methods that use probabilistic data structures and local Differential privacy for privacy preserving counting.

\subsection{Probabilistic data structures}
\label{subsec:prob_ds}

In the context of high-scale, low-latency and online data processing (virtually, there is never sufficient time nor resources), counts of items need to be calculated instantly as the data streams in, regardless of its scale.
Hence, instead of accurately keeping track of each item, we estimate the frequency using probabilistic data structures~\cite{Fan14,Mit05,Sch15}. 
Probabilistic data structures, such as Bloom filters and variants, sketches, and Cuckoo filters, have recently received much attention, as they are highly efficient for storing, processing, and computing~\cite{Dp17,Erl14,Fan14,Mit05,Sch15}.

\noindent
\textbf{Bloom filters}:
Bloom filters are bit vectors that initially contain $0$ in all the bit positions. $k$ independent hash functions $h_i(\cdot)$ (with $1 \le i \le k$) are used to hash-map an element $x$ by setting the corresponding bit positions in the Bloom filter $b$ to $1$ (i.e. $\forall_i~ b[h_i(x)] = 1)$. A Bloom filter allows a tunable false positive rate $fpr$ so that a query returns either “definitely not” (with no error), or “probably yes” (with probability $fpr$ of being wrong). The lower $fpr$ is, the better utility is, but the more space the filter requires.
The false positive probability for encoding $n$ elements into a Bloom filter of length $l$ bits using $k$ hash functions is $fpr = (1 - e^{-kn/l})^k$, which is controllable by tuning the parameters $k$ and $l$. 

\begin{definition}[Exact vs. fuzzy counting:]
Given a list of items $X$ and a query item $x$, exact counting of $x$ is $|\forall_{i \in X}~if~i == x|$, whereas fuzzy counting is $|\forall_{i \in X}~if~sim(i,x) \ge s_t|$, where $|\cdot|$ denotes the cardinality of the given set, $sim(\cdot)$ is a similarity function and $s_t$ is the minimum similarity threshold.
\end{definition}

The main feature of Bloom filter encoding that makes it applicable to efficient fuzzy counting is that it preserves the similarity/distance between an item and a queried item in the Bloom filter space (with a negligible utility loss)~\cite{Sch15,Vat16}.
For example, with string values the $q$-grams (sub-strings of length
$q$) of string values can be hash-mapped into the Bloom filter $b$ using $k$ independent hash
functions~\cite{Sch15}, 
while for numerical values, the neighbouring values (within a certain interval to allow fuzzy matching) of values can be hash-mapped into the Bloom filter~\cite{Vat16}. Figure~\ref{fig:BFs} illustrates an example of fuzzy counting of string and numerical values using Bloom filters~\cite{Sch15,Vat16}.

The similarity between Bloom filters can be calculated using a token-based similarity function, such as Jaccard, Dice, or Hamming~\cite{Vat13}. For example, Dice-coefficient is used in Figure~\ref{fig:BFs}, which is calculated as $2 \times \frac{\sum(b_1 \cap b_2)}{\sum(b_1)+\sum(b_2)}$, where $b_1$ and $b_2$ are the two Bloom filters. Due to collision of different elements being mapped to the same bit position that occurs during the hash-mapping (depending on the parameter setting), the Bloom filter-based matching might result in false positives. With appropriate parameter settings, Bloom filters have shown to be successful in providing high matching results~\cite{Sch15,Ran13,Vat16}.
However, Bloom filters don't allow storing the counts of items. 

Counting Bloom filter is a variation of conventional Bloom filters which allows storing counts as well as adding, deleting or updating of items. The filter comprises an array of $t$-bit buckets. When an item is added, the corresponding counters are incremented, and when it's removed, the counters are decremented.
Consequently, a counting Bloom filter takes $t$-times more space than a conventional Bloom filter, and it also has a scalability limit. Count query for an element $x$ from a counting Bloom filter $cbf$ returns $min(\forall_i~ cbf[h_i(x)])$.

\begin{figure*}[th]%
    \centering
    \includegraphics[width= 1.0\linewidth]{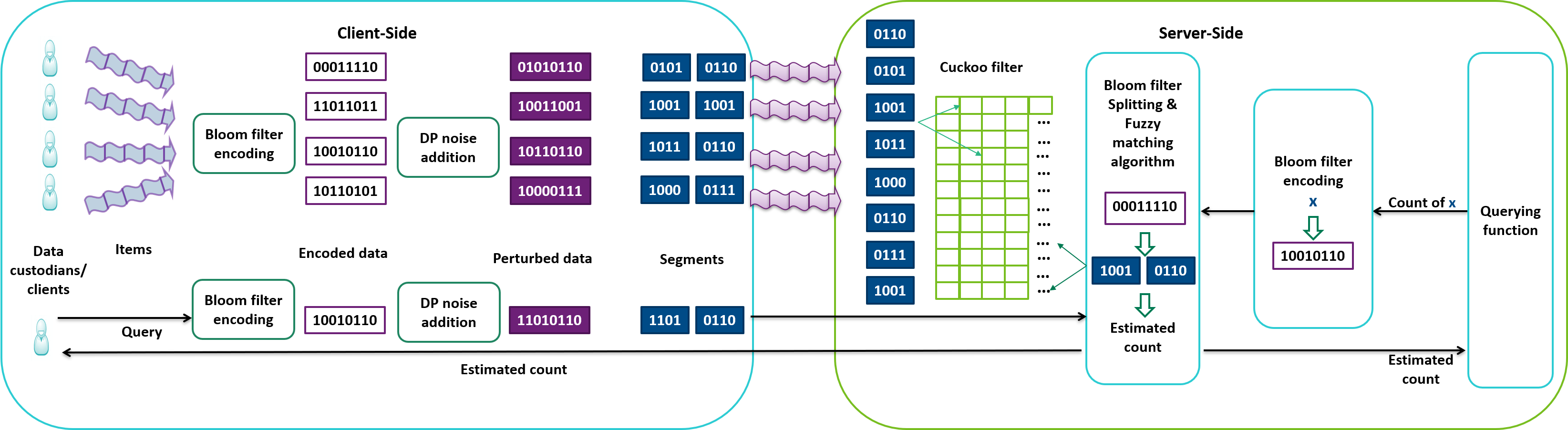}
    \caption{An overview of the proposed system for privacy preserving fuzzy online counting in stream data. 
    }
    \label{fig:sys_overview}
\end{figure*}

\noindent
\textbf{Sketches}:
A sketch is an array that consists of $D$ rows and $W$ cells in each row, initialized to 0, where $W$ and $D$ indicate the width and depth of the sketch, respectively. 
Given a pair of parameters $(\theta,\delta)$, the sketch parameters can be set as $W = \big \lceil e/\theta \big \rceil $ and $D = \big \lceil ln(1 / \delta) \big \rceil$, where $e$ is Euler’s number, and $\theta$ and $\delta$ mean that the error in answering a query is within a factor of $\theta$ with probability of $1 - \delta$.
In a count-min sketch, each element is hashed by randomly chosen pairwise independent hash functions and the corresponding cell values are incremented by the count of that element. The maximum probability of any two different elements being hash mapped to the same position is $1/W$ for each hash function, where $W$ is the width of the sketch or the number of cells. The false positive probability $fpr$ for count-min sketch is therefore, $fpr = [1 - (1 - 1/W)^n]^D$, where $n$ is the number of distinct elements stored in the sketch. 
Count query for an element $x$ from a sketch $M$ returns $min(\forall_i ~M[i][h_i(x)])$.
Several methods based on count-min sketch for the heavy-hitter problem were proposed~\cite{Cor12,Kara15,Mel16}. The heavy-hitter problem is about identifying frequent elements from data, for example IPs flooding in a network or over-consumed drugs to monitor disease outbreak. 

\noindent
\textbf{Cuckoo filter}:
The Cuckoo filter consists of a Cuckoo hash table that stores the ``fingerprints'' of items inserted. The fingerprint of an item is a bit string derived from the hash of that item. A Cuckoo hash table consists of an array of buckets where an item to be inserted is mapped to two possible buckets based on a hash function $h(\cdot)$. Each bucket can be configured to store a variable number of fingerprints. To insert an item $x$ into the Cuckoo filter, two indices need to be derived from the item based on hashing the item and its fingerprint:

\begin{align}
    &i1 = h(x), \\ \nonumber
    &i2 = i1 \oplus h(fp(x)),
    \label{eq:cf_buckets}
\end{align}

\noindent
where $h(\cdot)$ and $fp(\cdot)$ are the hash and fingerprint functions, respectively.

On obtaining these indices, the item's fingerprint ($f = fp(x)$) is inserted into one of the two possible buckets that correspond to the derived indices. $fp(\cdot)$ is used for space efficiency of the Cuckoo filter (depending on the fingerprint size)~\cite{Fan14}.
As the Cuckoo hash table begins to fill up, it encounters a situation where the two possible indices that an item can be inserted has been filled. In this case, items currently in the Cuckoo hash table are swapped (until the maximum number of swaps is met) to their alternative indices to free up space for inserting the new item. Querying the count of an item requires checking the two possible buckets for the fingerprint of the item. 

\subsection{State-of-the-art methods}
\label{subsec:SOTA}

The two state-of-the-art methods for probabilistic privacy preserving counting are:  
(1) Google's RAPPOR that uses Bloom filter and counting Bloom filter for collecting statistics about end-user software~\cite{Erl14}, and (2) Apple's algorithm presented in~\cite{Dp17} that uses Count-mean Sketches (CMS) to learn frequency of words, or web domains or emojis to e.g. determine the most popular emoji used by individuals or to identify high energy and memory usage in Apple's  web browser. Both of these methods only allow offline and exact count querying.
Cuckoo hashing/Cuckoo filter has been used in several private set intersection applications~\cite{Che17,Res18,Pin14,Pin15,Pin18} due to its significant space and time efficiency compared to other probabilistic data structures~\cite{Fan14}. However, Cuckoo filter has not been studied for privacy preserving counting. 

\subsection{Differential Privacy}
\label{subsec:DP}

Differential privacy~\cite{Dwo06,Dwo08,Dwo10} 
guarantees for each individual
in a dataset that any information that could be discovered about an individual with their data in
the dataset could also, with high probability, be discovered without their data in the
dataset. That is, the output of any query $f$ performed on dataset $x$ will be
indistinguishable from the output of the same query $f$ performed on dataset $y$, where $y$
differs from $x$ by at most one record (the record of any individual).

\begin{definition}
[Differential Privacy~\cite{Dwo06}] A randomized
function $\mathcal{A}$
(i.e. a function with a randomized component) is $\epsilon$-Differentially private if for all outputs
$y \subseteq Range(A)$ and for all data $x$, $x' \in \mathcal{D}^n$ such that $||x - x'||_1 \le 1$:

\begin{equation}
    Pr(\mathcal{A}(x) = y) \le e^{\epsilon} \times Pr(\mathcal{A}(x') = y).
\end{equation}

\end{definition}

The general differential privacy notion is defined for algorithms with input databases of size larger than 1. 
Local differential privacy (LDP) is a differential privacy model developed specifically to provide guarantees such that even if an adversary has access to the individual records/data in the dataset, the adversary is still unable to learn additional information about the individual from the individual data with high probability~\cite{Evf03}. It ensures differential privacy guarantees for each individual's inputs by processing (perturbing) the data locally on-device rather than processing in the central server. LDP has become the de-facto privacy standard around the world in recent years, with the technology companies Google and Apple implementing LDP in their latest operating systems and applications~\cite{Erl14,Dp17,Gre16}.  

Assume two adjacent streams of data where the two data streams differ at most by one record or item:
\begin{definition}
[Local Differential Privacy~\cite{Dwo06}] Let $\mathcal{A}: \mathcal{D} \rightarrow \mathcal{Y}$ be a randomized algorithm mapping a data entry in $\mathcal{D}$ to $\mathcal{Y}$. The algorithm $\mathcal{A}$ is $\epsilon$-local differentially private if for data entry $x, \neg x \in \mathcal{D}$ and all outputs $y \in \mathcal{Y}$, 
\begin{equation}
    -\epsilon \le \frac{Pr[\mathcal{A}(x) = y]}{Pr[\mathcal{A}(\neg x) = y]} \le \epsilon
\end{equation}
\end{definition}

RAPPOR uses the randomized response method to achieve $\epsilon$-LDP by flipping bits in the Bloom filters of encoded items sent by the data clients to the server for updates. Apple's count mean sketch aggregation proposes to flip the bits in vectors with probability $1/(e^{\epsilon/2} + 1)$ to meet $\epsilon$-LDP guarantees. 

\section{Methodology}
\label{sec:methodology}

In this section, we describe our proposed method for privacy preserving real-time and fuzzy counting in stream data using Cuckoo filter and Bloom filter.
The proposed system overview is shown in Figure~\ref{fig:sys_overview}.
At the client-side, new items are encoded, perturbed, and sent to the server, while at the server-side the (perturbed) items are stored in Cuckoo filter which allows fuzzy matching for count querying (similar) items. 
We first describe the space and time-efficient probabilistic data structure, Cuckoo filter, for enhanced efficiency for processing streams of data, and we make the filter adaptive to continuous flow of data. 
We then propose a novel local Differentially private algorithm for cuckoo hashing of encoded (into Bloom filters) items, and finally, we describe how Bloom filters can be combined with Cuckoo filters to allow fuzzy matching for querying of counts in the presence of data errors and variations.
Algorithm~1 outlines the insertion step, Algorithm~2 presents the steps of local Differential privacy mechanism, while Algorithm~3 outlines the querying step (lookup operation) using fuzzy matching. Algorithms~1 and~3 are used by the server while Algorithms~1 and 2 are used by the users/data custodian before sending the items to the server for insertion/update.

\subsection{Efficient and Adaptive Filter}
\label{subsec:adaptive_filter}

Cuckoo filter consists of an array of buckets where an item $x$ is hash-mapped by inserting the fingerprint ($fp(\cdot)$) of $x$ into one of two possible buckets based on a hash function $h(\cdot)$, which are $h(x)$ and $h(x) \oplus h(fp(x))$. 
In our proposed method, we insert Bloom filter encoding of the item into the Cuckoo filter (i.e. $x' = bf(x)$, where $bf(\cdot)$ is the Bloom filter encoding function as described in Section~\ref{sec:preliminaries}). Bloom filter encoding preserves the distances/similarities between items in the original space, and therefore allows 
fuzzy matching (as will be discussed in Section~\ref{subsec:fuzzy_matching}) 
as well as enables
adding noise to perturb data in the differential privacy mechanism without significant impact on utility loss (as will be described in Section~\ref{subsec:dp}). Our method hashes segments of Bloom filters into the Cuckoo filter (as shown in lines~1-3 in Algorithm~1), such that more similar items will have more number of similar Bloom filter segments hashed into the Cuckoo filter. 

\begin{table}[t]
\footnotesize
\addtolength{\tabcolsep}{-4pt}
\begin{tabular}{lll}
  \label{algo_insertion}
    ~ \\[0.5mm] \hline 
    \\[-2mm]
    \multicolumn{3}{l}{\textbf{Algorithm~1:} Insertion (Client-side and server-side)}
      \\[0.5mm] \hline
    ~ \\[-2mm]
    \multicolumn{3}{l}{\textbf{Input:}} \\
    \multicolumn{3}{l}{\footnotesize{- $x$: A data/element}} \\
    \multicolumn{3}{l}{\footnotesize{- $m$: Number of Bloom filter segments}} \\
    \multicolumn{3}{l}{\footnotesize{- $bf(\cdot)$: Bloom filter encoding function}} \\
    \multicolumn{3}{l}{\footnotesize{- $h(\cdot)$: Hash function}} \\
    \multicolumn{3}{l}{\footnotesize{- $fp(\cdot)$: Fingerprint function}} \\
    \multicolumn{3}{l}{\footnotesize{- $max\_num\_kicks$: Maximum number of kicks to relocate items}} \\
    \multicolumn{3}{l}{\textbf{Output:}} \\
    \multicolumn{3}{l}{\footnotesize{- $\mathbf{C}$: Updated Cuckoo filter}} \\[1mm]

    \multicolumn{2}{l}{\footnotesize{\textbf{Client-side}:}} & ~ \\
    \footnotesize{1:}& \footnotesize{$x' = bf(x)$} & // \footnotesize{Bloom filter encoding} \\
    \footnotesize{2:}& \footnotesize{$bf\_segs = x'.segment(m)$} & // \footnotesize{Split into m segments} \\
    \footnotesize{3:}&\footnotesize{$send\_to\_server(bf\_segs)$} & // \footnotesize{Send to server} \\[1mm]
    \multicolumn{2}{l}{\footnotesize{\textbf{Server-side}:}} & ~ \\
    \footnotesize{4:}& \footnotesize{\textbf{for} $bf\_seg \in bf\_segs$ \textbf{do}:} & \footnotesize{// Iterate segments} \\
    \footnotesize{5:}& \footnotesize{\hspace{2mm} $f = fp(bf\_seg)$} & \footnotesize{// fingerprint} \\
    \footnotesize{6:}& \footnotesize{\hspace{2mm} $i1 = h(bf\_seg)$} & \footnotesize{// First bucket index} \\
    \footnotesize{7:}& \footnotesize{\hspace{2mm} $i2 = i1 \oplus h(f)$} & \footnotesize{// Second bucket index} \\    
    \footnotesize{8:}& \footnotesize{\hspace{2mm} \textbf{if} $\mathbf{C}.bucket[i1]~not~full$ \textbf{then}} & ~ \\
    \footnotesize{9:}& \footnotesize{\hspace{2mm}\hspace{2mm} $\mathbf{C}.bucket[i1].add(f)$} & \footnotesize{// Add to first bucket} \\
    \footnotesize{10:}& \footnotesize{\hspace{2mm} \textbf{else if} $\mathbf{C}.bucket[i2]~not~full$ \textbf{then}} & ~ \\
    \footnotesize{11:}& \footnotesize{\hspace{2mm}\hspace{2mm} $\mathbf{C}.bucket[i2].add(f)$} & \footnotesize{// Add to second bucket} \\
    \footnotesize{12:}& \footnotesize{\hspace{2mm} \textbf{else}} & ~ \\
    \footnotesize{13:}& \footnotesize{\hspace{2mm}\hspace{2mm} $i=random(i1,i2)$}& ~ \\
    \footnotesize{14:}& \footnotesize{\hspace{2mm}\hspace{2mm} \textbf{for} $n=0; n \le max\_num\_kicks$ \textbf{do}}& \footnotesize{// Relocate items} \\
    \footnotesize{15:}& \footnotesize{\hspace{2mm}\hspace{2mm}\hspace{2mm} $f' = random(\mathbf{C}.bucket[i])$}& ~ \\
    \footnotesize{16:}& \footnotesize{\hspace{2mm}\hspace{2mm}\hspace{2mm} $f, f' = f', f$}& \footnotesize{// Swap values} \\
    \footnotesize{17:}& \footnotesize{\hspace{2mm}\hspace{2mm}\hspace{2mm} $i = i \oplus h(f)$}& ~ \\
    \footnotesize{18:}& \footnotesize{\hspace{2mm}\hspace{2mm}\hspace{2mm} \textbf{if} $\mathbf{C}.bucket[i]~not~full$ \textbf{then}}& ~  \\
    \footnotesize{19:}& \footnotesize{\hspace{2mm}\hspace{2mm}\hspace{2mm}\hspace{2mm} $\mathbf{C}.bucket[i].add(f)$} & \footnotesize{// Add f to bucket i} \\
    \footnotesize{20:}& \footnotesize{\hspace{2mm}\hspace{2mm}\hspace{2mm}\hspace{2mm} $break$} & \footnotesize{// Exit loop} \\
    \footnotesize{21:}& \footnotesize{\hspace{2mm}\hspace{2mm}\hspace{2mm} $insertion = fail$} & \footnotesize{// Insertion failed} \\
    \footnotesize{22:}& \footnotesize{\hspace{2mm} \textbf{if} $insertion == fail$ \textbf{then}} & ~ \\
    \footnotesize{23:}& \footnotesize{\hspace{2mm}\hspace{2mm} $\mathbf{C}.bucket\_size += \mathbf{C}.bucket\_size$}  & \footnotesize{// If full, adaptive filter} \\
    \footnotesize{24:}& \footnotesize{\hspace{2mm}\hspace{2mm} $\mathbf{C}.bucket[i1].add(f)$}  & \footnotesize{// Add to bucket i1} \\
    \footnotesize{25:}& \footnotesize{return $\mathbf{C}$} & \footnotesize{// Output $\mathbf{C}$} \\ 
      \hline 
  \end{tabular}
\end{table}

Cuckoo filters have been shown to be more efficient than Counting Bloom filters and Count-min Sketches in terms of space size and lookup operations~\cite{Fan14}. 
The space cost, in bits, of storing one item in the Cuckoo filter depends on the target false positive rate $fpr$ and is given by $(log_2(1/fpr)+2)/\alpha$, where $\alpha$ is the load factor of the filter which defines the maximum filter capacity. The false positive probability $fpr$ of Cuckoo filters where an element $x$ has the same fingerprint as another element $y$ and shares one of $x$'s buckets is $fpr = [2/O \times 1/2^F]$, where $O$ is the number of filled/occupied buckets and $F$ is the fingerprint size.
With increasing $O$, the $fpr$ increases.

To account for the impact of continuous flow of data on the false positive rate $fpr$, we make Cuckoo filter adaptive with regard to the size of the buckets in the filter. 
In contrast to Bloom filters and sketches, cuckoo filters can be adaptive to increasing volume of data insertions.
With Bloom filters or sketches (as used by RAPPOR~\cite{Erl14} and Apple's algorithm~\cite{Dp17}, respectively), it is not trivial to increase the size of these data structures to be adaptive, without re-hashing all the previously stored items in these data structures, and as a result the false positive rate increases with large number of items insertion and finally reaches to $1.0$. While with cuckoo filters, the bucket size can be incrementally increased depending on the percentage of occupied buckets.
If the filter is considered to be mostly occupied (full), then the length or size of the buckets is increased adaptively. Upon failure of an element insertion due to $O$ is large and exceeds $\alpha$ and therefore an empty space is not found until the maximum number of swaps is performed, we increment the filter's bucket size by the default size and continue with the insertion until $O$ remains less than $\alpha$ (lines~22-24 in Algorithm~1). 

The lookup operation for querying an item $x$ 
in the adaptive Cuckoo filter 
is still only in 2 buckets ($h(bf(x))$ and $h(bf(x)) \oplus h(f(bf(x)))$), however the number of lookup items in each bucket increases in the adaptive filter with the increasing number of records being inserted into the filter.

\begin{table}[t]
\footnotesize
\addtolength{\tabcolsep}{-6pt}
\begin{tabular}{lll} 
  \label{algo_early_map}
    ~ \\[0.5mm] \hline 
    \\[-2mm]
    \multicolumn{3}{l}{\textbf{Algorithm~2:} Noise addition for local differential privacy (Client-side)}
      \\[0.5mm] \hline
    ~ \\[-2mm]
    \multicolumn{3}{l}{\textbf{Input:}} \\
    \multicolumn{3}{l}{\footnotesize{- $x$: A data/element}} \\
    \multicolumn{3}{l}{\footnotesize{- $bf(\cdot)$: Bloom filter (BF) encoding function}} \\
    \multicolumn{3}{l}{\footnotesize{- $h(\cdot)$: Hash function}} \\
    \multicolumn{3}{l}{\footnotesize{- $\epsilon$: Privacy budget}} \\
    \multicolumn{3}{l}{\footnotesize{- $s_t$: Minimum similarity threshold}} \\
    \multicolumn{3}{l}{\footnotesize{- $p$: Probability to flip bits $p=\frac{1}{1+s.e^{\epsilon}}$ where $s=\frac{2^{l.(1-s_t)}.B}{2^l}$} (Thm~3.1)} \\
    \multicolumn{3}{l}{\footnotesize{- $F'$: Corresponding fingerprints for each bucket}} \\
    \multicolumn{3}{l}{\textbf{Output:}} \\
    \multicolumn{3}{l}{\footnotesize{- $V'$: Perturbed vector containing fingerprints}} \\[1mm]

    \footnotesize{1:}& \footnotesize{$x' = bf(x)$} & \footnotesize{//BF of $x$} \\
    \footnotesize{2:}& \footnotesize{$i1 = h(x')$} & \footnotesize{//First bucket} \\
    \footnotesize{3:}& \footnotesize{$V = \{-1 \in \mathbb{R}^B$\}} & \footnotesize{//Initialize a vector} \\
    \footnotesize{4:}& \footnotesize{$V[i1] = +1$} & ~ \\    
    \footnotesize{5:}& \footnotesize{$N \in \{-1, +1\}^B$, $Pr[n \in N = +1] = 1 - p$} & \footnotesize{//Noise vector} \\
    \footnotesize{6:}& \footnotesize{$V = [v_1n_1, \cdots, v_BN_B]$} & \footnotesize{//Flipping bits} \\
    \footnotesize{8:}& \footnotesize{\textbf{for} $v \in V$  \textbf{do}} & ~ \\
    \footnotesize{9:}& \footnotesize{\hspace{2mm}\textbf{if} $v == i1$} \textbf{then} & ~ \\
    \footnotesize{10:}& \footnotesize{\hspace{2mm}\hspace{2mm} $\tilde{x'}=generate\_similar\_bf(x', s_t)$} & \footnotesize{//similar BF with $s_t$} \\
    \footnotesize{11:}& \footnotesize{\hspace{2mm}\hspace{2mm} $V'.add(\tilde{x'})$} & \footnotesize{//Add similar BF $\tilde{x'}$} \\
    \footnotesize{12:}& \footnotesize{\hspace{2mm}\textbf{else}} & ~ \\
    \footnotesize{13:}& \footnotesize{\hspace{2mm} $x" = randomly\_choose\_from(F'[r])$} & \footnotesize{//Corresponding BF} \\
    \footnotesize{14:}& \footnotesize{\hspace{2mm} $V'.add(x")$}& \footnotesize{//Add artificial BFs} \\
    \footnotesize{15:}& \footnotesize{return $V'$} & \footnotesize{//Send $V'$ to server} \\ 
      \hline 
  \end{tabular}
\end{table}

\subsection{Privacy Guarantees}
\label{subsec:dp}

Similar to~\cite{Dp17}, the threat model of this research problem is as follows: encoded data (Bloom filters) of unique/new items from each client or custodian are continuously being sent to the (untrusted) server to be stored or updated in the cloud. Note that, in contrast to the problem addressed in~\cite{Erl14}, each client needs to report only unique items (i.e. if the item is not already reported by that client), and therefore the clients locally keep track of the unique items reported to the server. When a data consumer/query issuer (for example, a researcher
who wants to identify the most popular emojis used by many users)
queries the frequency counts of the items in the cloud from all users, the estimated counts are revealed. We assume the server is untrusted and the server knows the hash function used by data custodians.

The privacy preserving context for this problem requires that the server or the consumer of the protocol (query issuer) should not learn individual users' membership information of an item from the data, and an eavesdropper who gets access to the communication channel or to the server should not be able to learn the items about any individual users. However, the encoded data (Bloom filters) sent by custodians/clients can leak some information about individual users by performing a frequency attack~\cite{Chr18b} or dictionary attack using the hash encoding functions on some known values~\cite{Vat13}.

The domain space of the items is generally well-known and some of the items are considered to be highly sensitive. For example, if the application is counting the frequency of words entered by data custodians, some of the words such as related to cancer illness, are considered to be highly confidential for data custodians and therefore identifying a certain data custodian has typed words related to cancer illness raises serious privacy issues. 
Bloom filter encoding can be applied on the known (and sensitive/confidential) items from the problem domain to match the incoming Bloom filter encodings from different data custodians in order to infer the presence of a sensitive item, or they can be inferred using the frequency distribution of bit patterns (known as cryptanalysis attack~\cite{Chr18b}).

\subsubsection{\textbf{Local Differential privacy}}
\label{subsubsec:localdp}

Perturbing the bits in the Bloom filter itself incurs significant amount of utility loss. Even a single bit difference can correspond to completely different item values. 
Therefore, in our method clients do not add fake ``artificially-crafted'' bits in the Bloom filters, but they send possibly multiple Bloom filters even when there is a single true item. 

\smallskip
\noindent
\textbf{Pre-processing}:
As an offline and one-time pre-processing step, at the client-side a dictionary $F'$ containing corresponding Bloom filters for each of the $B$ bucket is generated. All the possible bit patterns of length $l$ with random 1's and 0's ($2^l$ in total) are generated and assigned to the two corresponding bucket indices ($i1$ and $i2$) based on the $h(\cdot)$ and $fp(\cdot)$ functions. 

\smallskip
\noindent
\textbf{Online processing}:
For every incoming item the real bucket index is first represented as a one-hot vector containing 1 at the position corresponding to the real bucket and -1 elsewhere. Randomised response noise mechanism is applied to this vector using our flipping probability $P_{flip} = \frac{1}{1 + s.e^{\epsilon}}$, where $s = \frac{2^{l.(1-s_t)}}{t}$, and $t = 2^l/B$,  derived from Theorem~\ref{th:dp} (as will be detailed below). Clients then send a Bloom filter corresponding to each bucket which is flipped to 1. For all the buckets other than the real bucket, a random Bloom filter chosen from the pre-computed dictionary for that bucket is sent. For the real bucket though, if the bit is not flipped to 0, a randomly chosen Bloom filter at distance of minimum similarity threshold $s_t$ and maximum $1.0$ is sent. 
The steps of our local Differential privacy mechanism are outlined in Algorithm~2:
\begin{enumerate}
    \item For the actual item x, the Bloom filter $x'= bf(x)$ and bucket position $i1 = h(x')$ are computed (lines 1-2 in Algorithm~2).
    \item A one-hot vector $V$ of size $B$ (where $B$ is the number of buckets in the cuckoo filter) is derived containing $1$ at position $i1$ and $-1$ everywhere else, and the randomised response noise mechanism is applied to $V$ using our flipping probability (derived from Theorem~\ref{th:dp}).
    \item A new vector $V'$ is constructed from $V$, where all the 1's are replaced by Bloom filters chosen as (lines 9-14):
    \begin{enumerate}
        \item  For all buckets other than $i1$, a random Bloom filter chosen from a pre-computed list $F'$ containing corresponding Bloom filters (using the $randomly\_choose\_from()$ function) is sent (lines~9-11).
        \item If $i1$ is not flipped (lines~12-14), a similar Bloom filter is generated (using the function $generate\_similar\_bf()$) with a distance $s_c$ chosen randomly from the range of $[s_t,1.0]$ and sent (where $s_t$ is the minimum similarity threshold to allow fuzzy matching). This Bloom filter is generated by randomly flipping $s_c \times l$ bits in the original Bloom filter, where $l$ is the length of Bloom filter. For example, in the running example shown in Fig.~\ref{fig:BFs}, the original Bloom filter `110110011' is perturbed as `110101011' if $s_c=0.9$ chosen from the range $[0.8, 1.0]$, as $l \times s_c = 9 \times 0.9 = 2$ random bits ($5^{th}$ and $6^{th}$ in this example) need to be flipped.
    
      \end{enumerate}
    \item The vector $V'$ is sent to the server.
\end{enumerate}

In the following we provide the formal proof of Algorithm~2 being $\epsilon$-local Differentially private for the adjacency of two neighbouring streams of data that differ by one item ($x_1$) which is present in one stream and absent in the other. We also provide the proof for the adjacency items being different items ($x_1$ in one stream and $x_2$ in the other) in Appendix~\ref{app:proof}. 

\begin{theorem}[Differential privacy of Algorithm 2]
\label{th:dp}
Algorithm 2 is $\epsilon$-local differentially private.
\end{theorem}

\begin{proof}
For an arbitrary output $\tilde{V'}$, the ratio of probabilities of observing the same output given a user reported item $x_1$ or not is: 

\begin{footnotesize}
\begin{equation}
    \frac{Pr[PPCF(x_1, \epsilon) = \tilde{V'}]}{Pr[PPCF(\neg x_1, \epsilon) = \tilde{V'}]} = \frac{Pr[[b_{1,1}, b_{2,1}, \cdots, b_{u,1}]=\tilde{V'}]}{Pr[[b_{1,2}, b_{2,2}, \cdots, b_{u,2}] = \tilde{V'}]},
\label{eq:1}
\end{equation}
\end{footnotesize}
where $u$ is the number of bits flipped to +1 in the vector $V$, and $b_{i,1}$ denotes the $i^{th}$ Bloom filter sent for item $x_1$ and $b_{i,2}$ denotes the $i^{th}$ Bloom filter sent for item that is not $x_1$. 
At maximum, only one $b_{i,j}$ (where $i=1$ or $i=2$) in each of the set $V'_{j}$ can correspond to the real item $x_j$, while others correspond to artificially crafted Bloom filters.

The probability of $b_{i,j}$ corresponding to artificial items being the same in two output sets $V'_1$ and $V'_2$ (sent for two items' $x_1$ and $\neg x_1$ insertion) is same. Without loss of generality, let's assume except the first Bloom filter (i.e. $b_{1,j}$), all the other Bloom filters are artificial. The above ratio is then maximized when the Bloom filters of the real items is present in one of the output sets (say $V'_1$) 
(i.e. $b_{1,1}$ is the Bloom filter encoding of real item $x_1$ and $b_{1,2}$ is the Bloom filter encoding of an artificial item). 

We denote by $P_{11}$ the probability that a reported Bloom filter $b_{1,j} \in V'_j$ is the encoding of the real item (say $x_1$) and is equal to $\tilde{v'}$, and by $P_{01}$ the probability that $b_{1,j} \in V'_j$ is an artificial item and is equal to $\tilde{v'}$.

\begin{footnotesize}
\begin{equation}
\begin{multlined}
    \frac{Pr[PPCF(x_1,\epsilon)= \tilde{V'}]}{Pr[PPCF(\neg x_1,\epsilon)=\tilde{V'}]} = \frac{P_{11} \times P_{01} \cdots \times P_{01}}{P_{11} \times P_{01} \cdots \times P_{01}} 
    = \frac{P_{11}}{P_{01}}
\label{eq:2}
\end{multlined}
\end{equation}
\end{footnotesize}

We first calculate $P_{11}$. This occurs when item $x_1$'s bucket index is not flipped in the vector $V$, i.e. $h(bf(x_1)) = +1$ and the random $s_t$-close Bloom filter chosen for this bucket is the same as $bf(x_1)$.

\begin{equation}
\begin{multlined}
    P_{11} = (1 - P_{flip}) \times \frac{1}{2^{l \times (1 - s_t)}} 
\label{eq:3}
\end{multlined}
\end{equation}

When $s_t == 1.0$,  $P_{11}$ depends only on the flip probability $P_{flip}$.

Next, let's calculate $P_{01}$. This can occur due to two reasons:
\begin{enumerate}
    \item An artificial bucket index gets flipped and $\tilde{v'}$ is chosen from the pre-filled set of $t$ Bloom filters for that bucket. 
    \item $\tilde{v'}$ gets chosen due to its $s_t$-closeness to a real item's Bloom filter.
\end{enumerate}

The probability of randomly choosing one Bloom filter from the corresponding Bloom filters for each bucket is $1/t$, where $t = 2^l/B$ assuming the Bloom filters/bit vectors are uniformly distributed across all buckets such that each bucket is assigned with $2^l/B$ corresponding Bloom filters. Then,

\begin{equation}
\begin{multlined}
P_{01} = [P_{flip} \times \frac{1}{t}] + [(1 - P_{flip}) \times \frac{1}{2^l}].
\label{eq:4}
\end{multlined}
\end{equation}

Hence, for the maximum ration to be bounded requires:

\begin{equation}
\begin{multlined}
    e^{-\epsilon} \le \frac{P_{11}}{P_{01}} \le e^{\epsilon} \\
    e^{-\epsilon} \le \frac{(1 - P_{flip}) \times \frac{1}{2^{l \times (1 - s_t)}}}{[P_{flip} \times \frac{1}{t}] + [(1 - P_{flip}) \times \frac{1}{2^l}]} \le e^{\epsilon} \\
\label{eq:5}
\end{multlined}
\end{equation}

Ignoring the $[(1 - P_{flip}) \times \frac{1}{2^l}]$ term, we bound the above ratio for  $P_{flip} \geq \frac{1}{1+s. e^{\epsilon}}$, where $s=\frac{2^{l(1-s_t)}}{t}$.

\begin{equation}
\begin{multlined}
    -\epsilon \le log(\frac{Pr[PPCF(x_1, \epsilon) = \tilde{V'}]}{Pr[PPCF(\neg x_1, \epsilon) = \tilde{V'}]} \le \epsilon \\
\label{eq:6}
\end{multlined}
\end{equation}

\end{proof}

\begin{table}[t]
\footnotesize
\addtolength{\tabcolsep}{-3pt}
\begin{tabular}{lll} 
  \label{algo_early_map}
    ~ \\[0.5mm] \hline 
    \\[-2mm]
    \multicolumn{3}{l}{\textbf{Algorithm~3:} Count querying with fuzzy matching (server-side)}
      \\[0.5mm] \hline
    ~ \\[-2mm]
    \multicolumn{3}{l}{\textbf{Input:}} \\
    \multicolumn{3}{l}{\footnotesize{- $x$: Query data/element}} \\
    \multicolumn{3}{l}{\footnotesize{- $m$: Number of Bloom filter segments}} \\
    \multicolumn{3}{l}{\footnotesize{- $bf(\cdot)$: Bloom filter encoding function}} \\
    \multicolumn{3}{l}{\footnotesize{- $h(\cdot)$: Hash function}} \\
    \multicolumn{3}{l}{\footnotesize{- $fp(\cdot)$: Fingerprint function}} \\
    \multicolumn{3}{l}{\footnotesize{- $s_t$: Minimum similarity threshold}} \\
    \multicolumn{3}{l}{\textbf{Output:}} \\
    \multicolumn{3}{l}{\footnotesize{- $c$: Estimated count of $x$}} \\[1mm]

    \footnotesize{1:}& \footnotesize{$x' = bf(x)$} & \footnotesize{//Bloom filter} \\
    \footnotesize{2:}& \footnotesize{$bf\_segs = x'.segment(m) $} & \footnotesize{//Segments} \\
    \footnotesize{3:}& \footnotesize{$C = [~]$} & ~ \\
    
    \footnotesize{4:}& \footnotesize{\textbf{for} $bf\_seg \in bf\_segs$ \textbf{do}} & \footnotesize{//Iterate segments} \\
    \footnotesize{5:}& \footnotesize{\hspace{2mm} $f = fp(bf\_seg)$} & \footnotesize{//Fingerprint} \\
    \footnotesize{6:}& \footnotesize{\hspace{2mm} $i1 = h(bf\_seg)$} & \footnotesize{//First bucket} \\
    \footnotesize{7:}& \footnotesize{\hspace{2mm} $i2 = i1 \oplus h(f)$} & \footnotesize{//Second bucket} \\    
    \footnotesize{8:}& \footnotesize{\hspace{2mm} \textbf{if} $\mathbf{C}.bucket[i1]~or$} &  ~ \\
    \footnotesize{}& \footnotesize{\hspace{2mm}\hspace{2mm}\hspace{2mm} $\mathbf{C}.bucket[i2]~has~f$ \textbf{then}} &  ~ \\
    \footnotesize{9:}& \footnotesize{\hspace{2mm}\hspace{2mm} $C.add(count(f))$} & \footnotesize{//Add $f$'s count to $C$} \\
    \footnotesize{10:}& \footnotesize{\hspace{2mm} \textbf{else}} &  ~ \\
    \footnotesize{11:}& \footnotesize{\hspace{2mm}\hspace{2mm} $C.add(0)$} & \footnotesize{//Add 0 count to $C$} \\
    \footnotesize{12:}& \footnotesize{\hspace{2mm} $0$-count\_segs = $[j \in C~if~j == 0]$} & \footnotesize{// $0$-count segments} \\
    \footnotesize{13:}& \footnotesize{\hspace{2mm} $sim_{max} = (m - |0$-count\_segs$|)/m $} & \footnotesize{//Max similarity} \\
    \footnotesize{14:}& \footnotesize{\hspace{2mm} \textbf{if} $sim_{max} \ge s_t$ \textbf{then}}& ~ \\
    \footnotesize{15:}& \footnotesize{\hspace{2mm}\hspace{2mm} $c = min([j~for~j \in C~if~j > 0])$}& ~ \\
    \footnotesize{16:}& \footnotesize{\hspace{2mm} \textbf{else}} & ~ \\
    \footnotesize{17:}& \footnotesize{\hspace{2mm}\hspace{2mm} $c = 0$}& ~ \\
    \footnotesize{18:}& \footnotesize{\hspace{2mm} return $c$} &  ~ \\ 
      \hline 
  \end{tabular}
\end{table}

\subsection{Fuzzy Counting}
\label{subsec:fuzzy_matching}

Real-world data often contains variations and errors (e.g. typos or misspellings in words), and therefore counting of items that differ by small typos or variations from the queried item is important in many real applications (e.g. word auto-correction application that counts words entered by different users even with small typos and errors). Further, range queries are commonly used in financial applications which require fuzzy counting of similar items (e.g. similar salary range).

In order to allow fuzzy matching, the data custodians first encode the item ($x$) into a Bloom filter, add noise in terms of artificially crafted Bloom filters (output set $V'$ from Algorithm~2),
and then split each of the Bloom filters in $V'$ (both real and artificial) into $m$ segments (lines~1-2 in Algorithm~1). 
The larger the value for $m$ is, the more accuracy the fuzzy matching will be. In Fig.~\ref{fig:BFs}~(right), an example fuzzy counting for the two example pairs of (item, query) values in Fig.~\ref{fig:BFs}~(left) and~(middle) using $m=3$ is illustrated.
The value for $m$ determines the degree of fault-tolerance of the method to data errors and variations. A smaller value of $m$ is sufficient for a clean data, however if the data is assumed to be largely dirty or for range queries with larger range, we need a large value of $m$ to allow effective fuzzy matching (please see Theorem~\ref{th:error}).

Then each of the $m$ Bloom filter segments are sent to the server (line~3 in Algorithm~1) to be inserted into the Cuckoo filter. If the Bloom filter segments of a querying item matches with a certain number of Bloom filter segments $s_m$ in the Cuckoo filter (depending on the similarity threshold parameter, $s_t$, given by the query issuer), where $s_m = m \times s_t$, then the estimated count of the most similar item to the queried item will be returned to the query issuer. In the example illustrated in Fig.~\ref{fig:BFs}, if $s_t=0.65$, then the items need to match in at least $s_m=2$ Bloom filter segments of the queried item in order to be counted. If the similarity threshold is set to $s_t = 1.0$, then it allows only exact matching, i.e. the estimated count of the item that is matching exactly the same as the queried item (in all $m$ segments) will be returned.

In contrast to the original Cuckoo hashing approach that stores fingerprints of items, our approach stores fingerprints of Bloom filter segments of items (lines~4-24 in Algorithm~1). Bloom filter segments preserve distance and therefore allow fuzzy matching (as discussed in Section~\ref{subsec:prob_ds}), whereas fingerprints of items do not, i.e. a single character difference (in string data) or number difference (in numerical data) returns completely different fingerprints. 
Note that the Bloom filters in the output $V'$ from Algorithm~2 are differentially private using our proposed noise addition algorithm. Bloom filter segmentation is a post-processing function that enables fuzzy matching.
Moreover, the Bloom filter segments of the Bloom filters (real and artificial) in $V'$ can be shuffled to amplify the privacy guarantees, as similar to the Encode, Shuffle, and Analyze mechanism proposed by Google~\cite{Bit17}.

As shown in Algorithm~3, the querying function returns the count value of the closest value to the querying value. For example, if the query value is $q=5000$, it will return the count/frequency of $5000$ if available in the filter (exact matching), otherwise it will return the count/frequency of a value $v$ that is the closest to $q$ and has a similarity above the similarity threshold $s_t$ (i.e. $sim(v,q) \ge s_t$ and that has the highest similarity with $q$). The algorithm first retrieves the count values (frequency/number of occurrences as calculated by the $count(\cdot)$ function) for each of the Bloom filter segments of the query value $q$ (lines~3-11 in Algorithm~3) and calculates how many segments have a count of $0$, i.e. not found in the filter (line~12).

Using this, it computes the maximum similarity of the closest value (line~13, where $|\cdot|$ denotes the cardinality of a given set) and if this similarity is above the similarity threshold $s_t$ (line~14) then the minimum count of the segments with non-zero counts is returned as the count value of the closest value of $q$ (lines~15-18).

\subsection{Utility analysis}
\label{subsec:utility_analysis}

We next analyze the utility loss with our proposed method.

\begin{theorem}[Count estimation error bound]
\label{th:error}
The bound of the estimated count $c'$ of an element $x$ using our method is given as.

\begin{equation}
\begin{multlined}
    c - c \times max(\frac{1}{1+s.e^{\epsilon}},  \frac{s.e^{\epsilon} \times (1 - s_t)^{m \times s_t}}{1+s.e^{\epsilon}} 
    ) \le c' \\
    \le c + [\frac{n \times (B-1)  \times 2^{l.(1-s_t)}}{(1+s.e^{\epsilon}) \times 2^l}], \\
\label{eq:5}
\end{multlined}
\end{equation}

where $c$ is the actual count of $x$, $n$ is the number of records inserted into the Cuckoo filter, $l$ is the length of Bloom filters, $m$ is the number of Bloom filter segments, $B$ is the number of buckets, and $s = \frac{2^{l.(1-s_t)}.B}{2^l}$ (see Theorem~\ref{th:dp})
\end{theorem}

\begin{proof}
The deviation of the estimated count $c'$ of an element $x$ from its actual count $c$ can occur due to false negatives as well as false positives leading to lower and higher estimated counts than the true counts: 

\textbf{Lower bound}: 
There are two cases associated with false negatives for an item $x$ (Bloom filter of $x$ is $x'$) resulting from noise addition to meet differential privacy guarantees.

\textbf{Case 1}: the bit $h(x')$ is flipped in $V$. This probability is $\frac{1}{1+s.e^{\epsilon}}$, i.e. from $c$ records of item $x$, $c \times \frac{1}{1+s.e^{\epsilon}}$ records would have been placed in wrong buckets in the Cuckoo filter (not in one of the buckets $h(x')$ or $h(x') \oplus h(fp(x'))$) and therefore resulting in false negatives.

\textbf{Case 2}: the bit $h(x')$ is not flipped in $V$, however the similar Bloom filter (similar to $x'$) chosen for bucket $h(x')$ does not match to $x'$ in the minimum number of segments out of $m$ segments according to the threshold $s_t$.
The probability that the bit $h(x')$ in the vector is not flipped to $-1$ is $\frac{s.e^{\epsilon}}{1+s.e^{\epsilon}}$. The minimum number of segments that need to be matched is $s_m = m \times s_t$. False positives occur if at least one bit in each of these segments gets flipped. The probability of a bit in $x'$ being flipped in $\tilde{x'}$ is $l(1-s_t)/l = 1-s_t$. 
Hence, the probability that at least 1 bit in each of the $s_m$ segments being flipped is $(1-s_t)^{m \times s_t}$.

Therefore, the lower bound of the estimated count $c'$ is $c - c~\times $ the maximum of the two cases, i.e. $max(\frac{1}{1+s.e^{\epsilon}}, \frac{s.e^{\epsilon} \times (1 - s_t)^{m \times s_t})}{1+s.e^{\epsilon}})$. 
For example, if $l=20$, $\epsilon = 6$, $s_t=0.7$, $m=4$, and $B=10000$, 
then $s=0.61$, and therefore $P_{flip} = \frac{1}{1+s.e^{\epsilon}} = 0.004$. The lower bound of the estimated frequency count for an item of true frequency count of $c=100$ is $c - c \times max(0.004, \frac{0.61 \times 403.4288 \times (0.3)^4)}{1 + (0.061 \times 403.4288)} ) = c - c \times max(0.004, 0.008) = 100 - 100 \times 0.008 = 99.19$.
With a larger number of segments $m=5$, the lower count becomes $c - c \times max(0.004, 0.0002) = 99.75$, and with a smaller $m=2$, the lower count becomes $c - c \times max(0.004,0.089) = 91.03$.

\textbf{Upper bound}: 
On the other hand, false positives could occur due to noise addition (where bits with -1 in $V$ are flipped to 1) resulting in insertion of corresponding (artificial) fingerprints into buckets and collision of fingerprints of different elements. The likelihood of any record being falsely inserted into the given bucket is $\frac{1}{1+s.e^{\epsilon}}$.  
Given $n$ is the number of elements/records that are inserted into the Cuckoo filter, $n \times (B-1) \times \frac{1}{1+s.e^{\epsilon}}$ records can be falsely inserted into the (artificial) buckets. Among these, there are $2^{l(1-s_t)}$ possible bit patterns that could match with the given real item's Bloom filter. The probability of choosing one of these potentially matching bit patterns from the artificial buckets is $\frac{2^{l(1-s_t)}/B}{2^l/B}$. Therefore, the upper bound of $c'$ is $c + [\frac{n \times (B-1)}{1+s.e^{\epsilon}} \times \frac{2^{l.(1-S_t)}}{2^l}]$.
With the previous example, if the number of records inserted so far is $n=1000$, then the upper bound of the estimated frequency count of an item whose real frequency count is $c=100$ is $100 + [\frac{1000 \times 9999}{1+(0.61 \times 403.4288)} \times \frac{2^{6}}{2^{20}}] = 100 + 2.47 = 102.47$. 
\end{proof}

\subsection{Communication and computation overhead}

Inserting an item into the Cuckoo filter requires sending $m \times B \times \frac{1}{1+s.e^{\epsilon}}$ Bloom filter segments of length $l/m$ bits, where $s = \frac{B.2^{l(1-s_t)}}{2^l}$ (see Theorem \ref{th:dp}). The communication complexity of insertion step therefore depends on the number of segments $m$, number of buckets $B$, and the length of Bloom filters $l$. Our method has higher communication overhead for the insertion step than RAPPOR and Apple's CMS algorithm which require only sending a vector of size $l$ bits to the server followed by vector operation between two vectors (between the received vector and the aggregated vector stored in the server) for inserting an item. 
However, we parallelize the insertion step to reduce the computation complexity by having multiple cuckoo filter data structures in the server and dividing the segments and inserting them into different structures. The querying function therefore needs to be applied on each of these data structures and the counts are summed to get the total frequency count of a queried item. This is conceptually similar to using different cohorts in RAPPOR~\cite{Erl14}.
Hence, the server-side querying of our method requires $2 \times m$ lookup operations at each of the cuckoo filters (in parallel) for every queried item and therefore querying is efficient than the RAPPOR and Apple's algorithm.  

\medskip
In summary, our method provides significantly higher utility guarantees and querying efficiency with similar privacy guarantees at the cost of communication overhead for inserting items. 

\section{Experimental Evaluation}
\label{sec:experiments}

In this section we present and discuss the results of experimental study of our proposed method.

\subsection{Datasets and Synthetic Dataset Generation}
\label{subsec:datasets}

We used two datasets in our experiments. 
The first one is words dataset that contains a list of words from the complete work of Shakespeare~\footnote{available from https://data.world/tronovan/shakespeare-word-frequencies}. The number of unique words in this dataset is 23,113. We duplicated these words to generate 1,750,000 records and split across 20 parties and the total number of records is 583,456. The total number of records is randomly divided into 10,000 disjoint sets for simulating 10,000 parties.

We further generated a synthetic dataset based on this words dataset for fuzzy counting experiments. For each of the unique values in the dataset, we created synthetic duplicate records and/or queries by replacing some of the values $v$ with similar values $v'$ according to the similarity threshold used ($s_t =0.8$).
Specifically, the synthetic dataset is generated by replacing 50\% of the count of each value with a similar value according to $s_t$.
To generate synthetic values $v'$ that are similar to $v$ (i.e., $sim(v,v') \ge s_t$), we modified $v$ by including character edits (inserts with $0.3$ probability, deletes with $0.3$ probability, and swaps with $0.4$ probability). 
We use these synthetically generated records for querying in order to evaluate and compare the performance of counting algorithms in the presence of data errors and variations.

True count for exact matching of a value in this dataset is the actual frequency/count of that value, while true count for fuzzy matching of a value is the count of all values similar to the value according to a similarity metric and similarity threshold. We use the $q$-gram-based Dice-coefficient metric for calculating the similarity between (unencoded) string data~\cite{Vat16,Chr12}. Estimated count of a value is the count value returned by the privacy preserving counting algorithm (Algorithm~2) when querying for that value.

\subsection{Baseline Methods and Parameter Setting}
\label{subsec:baseline}

We compare our proposed method with two state-of-the-art privacy preserving counting methods proposed by Google (RAPPOR)~\cite{Erl14} and Apple (CMS)~\cite{Dp17} as these are closely related to our work. Similar to our approach, these two methods use probabilistic data structures and local Differential privacy for privacy preserving counting (however, not specifically developed for real-time counting in streaming data applications, as described in detail in Section~\ref{sec:preliminaries}). 

\begin{figure*}[ht!]
\centering
 \includegraphics[width=0.38\textwidth]{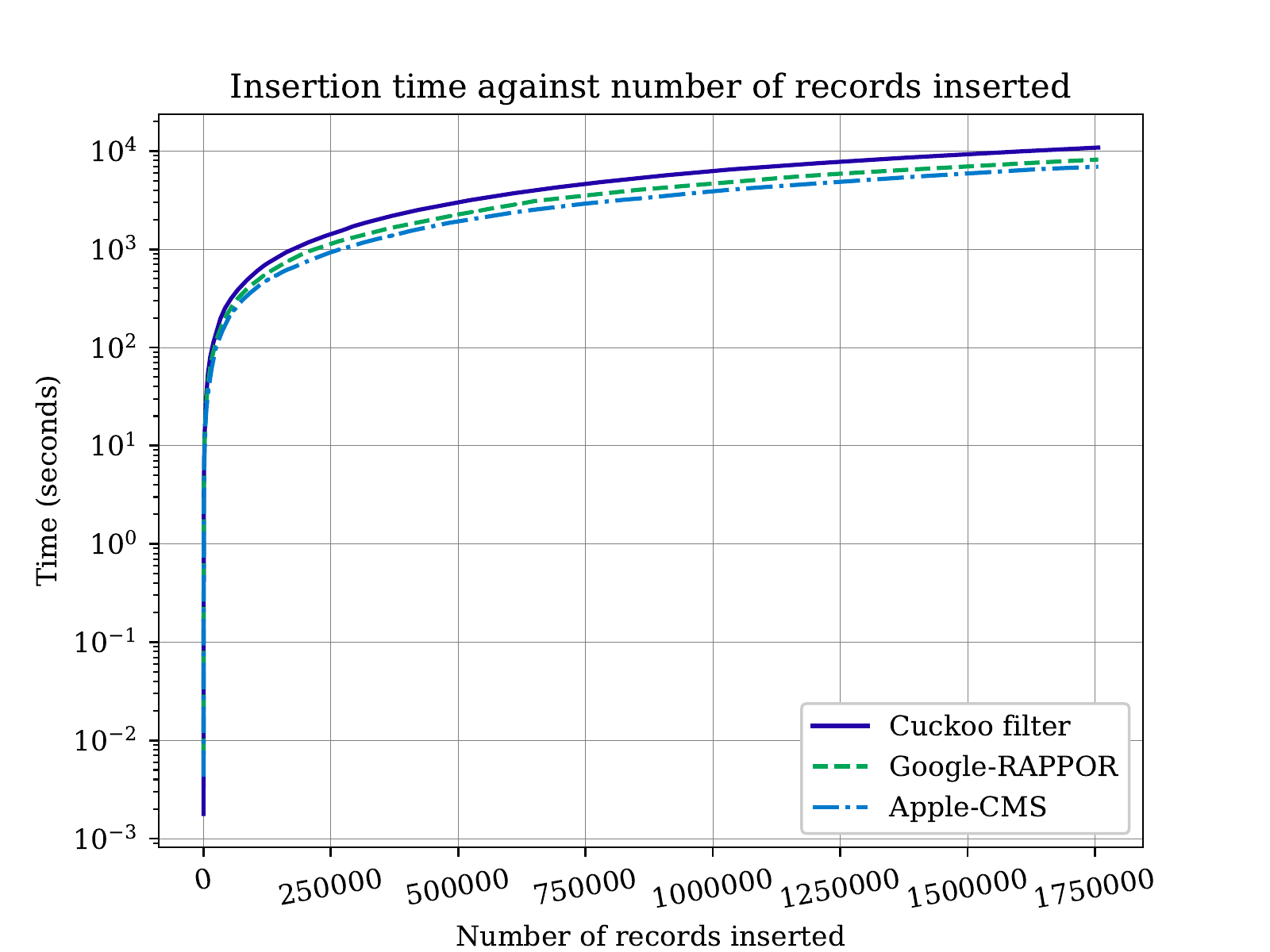}
 \includegraphics[width=0.38\textwidth]{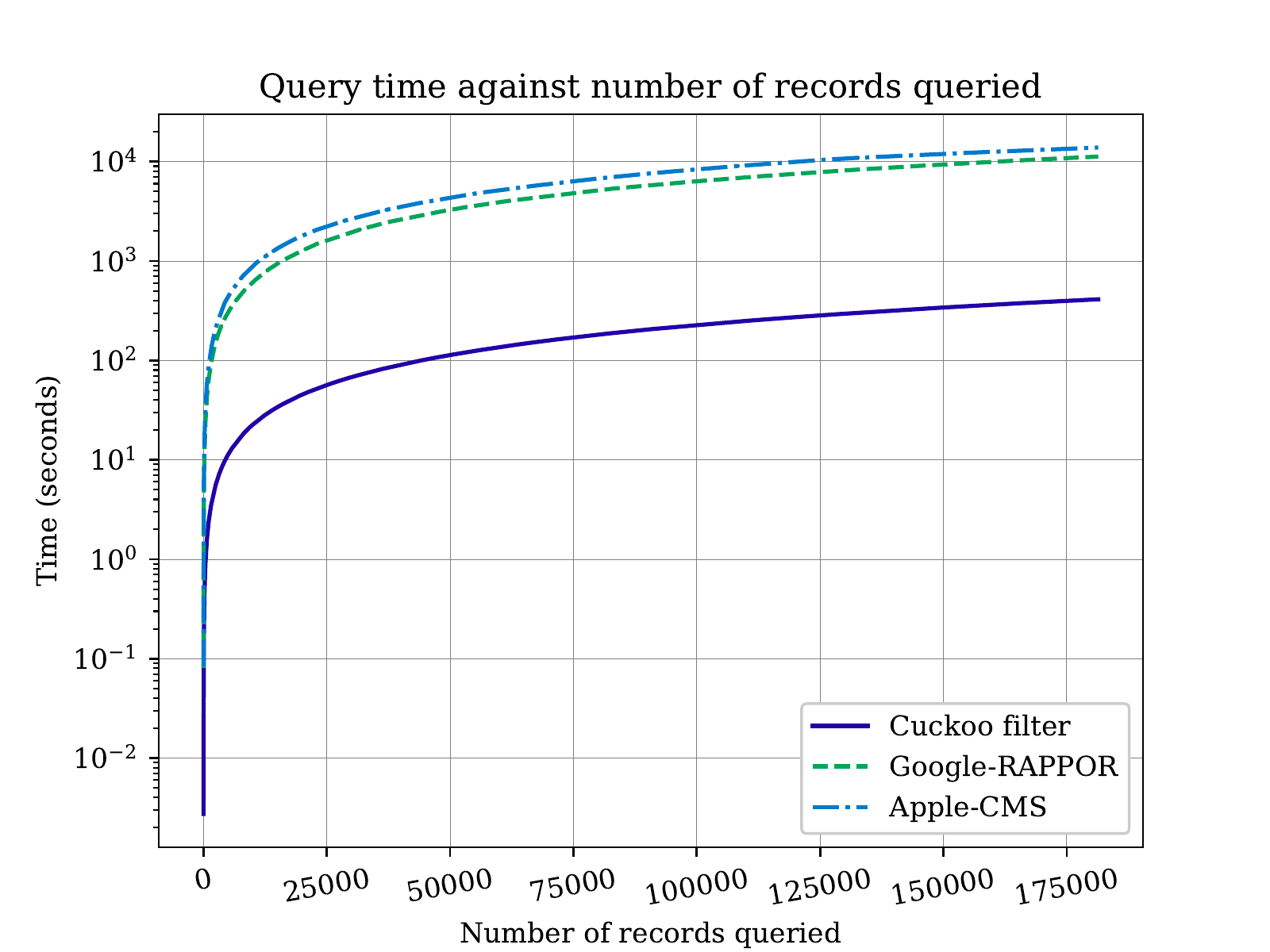} 
  \caption{\small{Comparison of efficiency results in terms of (a) 
  insertion time (CDF) and (b) querying time (CDF) for all approaches with exact counting (i.e $m=1$ with Cuckoo filter approach) on the words datasets.
  }
    }
\label{fig:comp_efficiency}
\end{figure*}

\begin{figure*}[ht!]
\centering
 \includegraphics[width=0.38\textwidth]{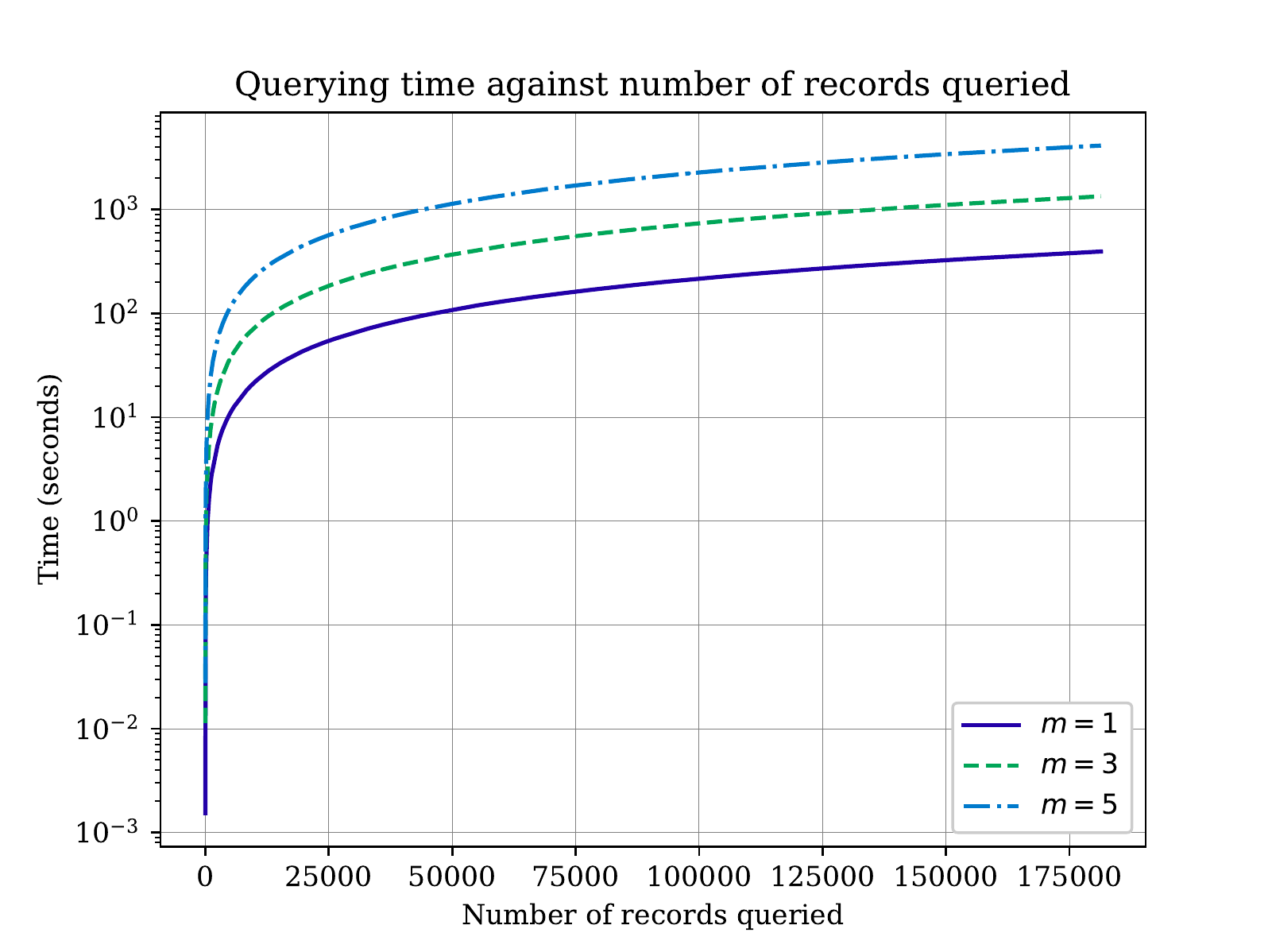} 
 \includegraphics[width=0.38\textwidth]{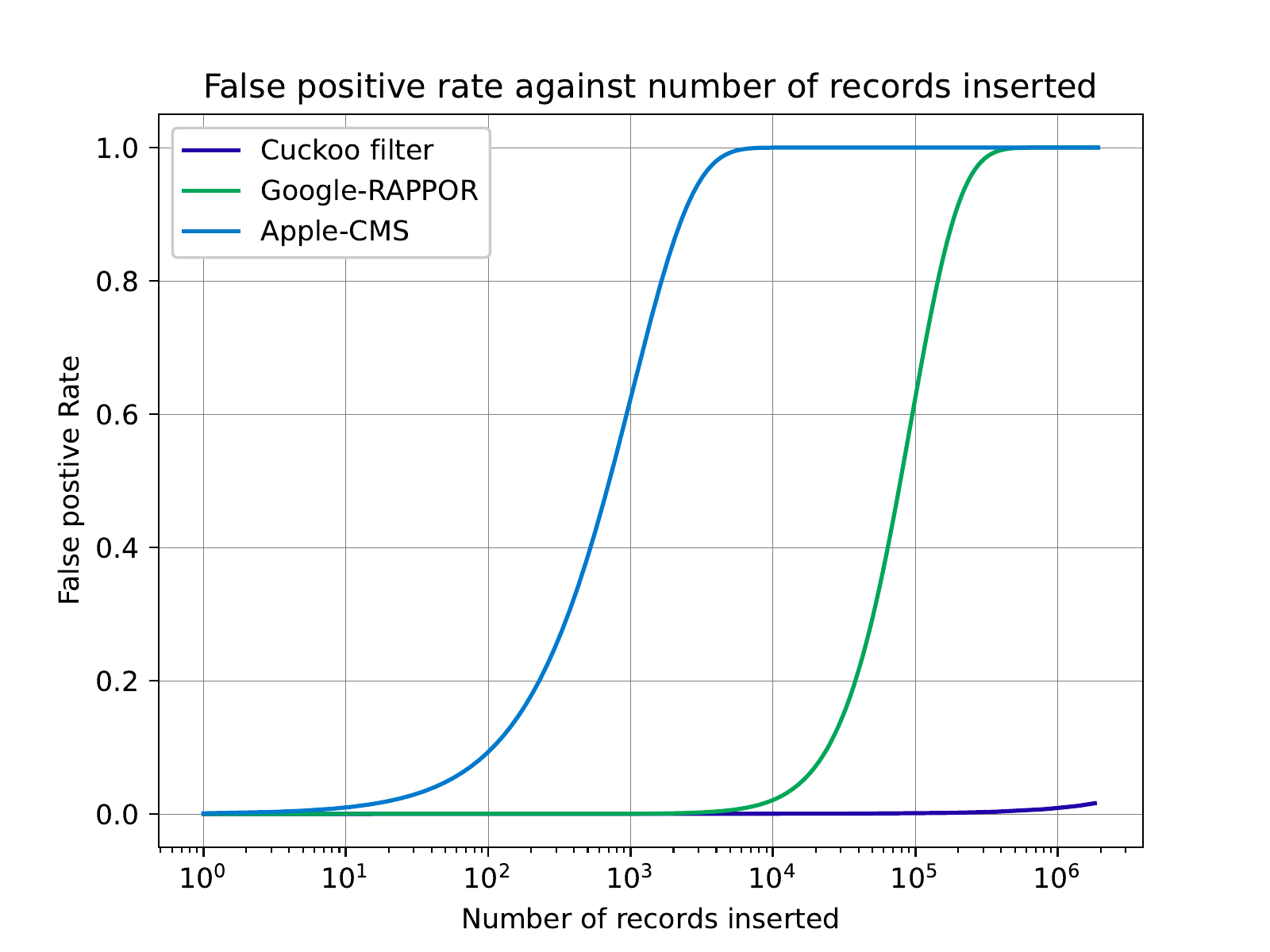}
  \caption{\small{Comparison of (a) fuzzy matching efficiency in terms of querying time (CDF) for different values of $m$ and (b) the impact of increasing volume of records insertion on the false positive rates for all approaches on the synthetic words datasets.
  }
    }
\label{fig:comp_efficiency_fuzzy}
\end{figure*}

We implemented both our proposed approach and the two competing 
baseline approaches in Python 3.5.2, and ran all experiments on a server with four 2-core 64-bit Intel Core I7 2.6 GHz CPUs, 8 GBytes of memory and running Ubuntu 16.04. Please note that the RAPPOR source code (implemented using R combined with Python) is publicly available~\footnote{available from https://github.com/google/rappor}. In order to perform a fair comparison, we implemented all three approaches in Python (using Python libraries such as numpy, bitarray, and sklearn). The programs and test datasets are available from the authors.

Default parameter setting of our approach is $\epsilon = 6$, capacity of Cuckoo filter $B=10000$, default bucket size $b=4$, maximum number of kicks $500$, similarity threshold $s_t = 0.8$, length of Bloom filters $l=30$, number of segments $m=5$, and number of hash functions $k = 2$, as it gave the best results when validating with a grid search parameter tuning method. We also evaluated our method against different $\epsilon$, in the range of $\epsilon = [2,4,6,8,10]$, and $m = [1,3,5]$.

Parameter settings of the state-of-the-art methods are chosen in a similar range (adjusted according to the number of records used) as used in the original approaches. For  Google-RAPPOR, we used the number of hash functions $k=2$, length of Bloom filters $l=1000$, number of cohorts $m=32$, and noise parameters $f=0.5$, $p=0.5$, and $q=0.75$. For Apple-CMS, we set $\epsilon = 8$, length of vectors $l=1024$, and the number of hash functions $k=20000$. 

\subsection{Evaluation metrics}
\label{subsec:metrics}

We evaluated our method and compared with Google-RAPPOR and Apple-CMS in terms of insertion and querying efficiency as well as accuracy of querying results. Computational efficiency is measured using insertion time (in seconds) and querying time (in seconds).  
The accuracy of count estimation is measured as the absolute error that calculates the variance/difference between true count and estimated count (i.e. the estimation error). We also evaluated the false positive rate of the probabilistic data structures against number of records inserted to evaluate the impact of continuous data insertion on the utility. Privacy is measured using the privacy budget $\epsilon$ required for providing Differential privacy guarantees.

\begin{figure*}[ht!]
\centering
 \includegraphics[width=0.38\textwidth]{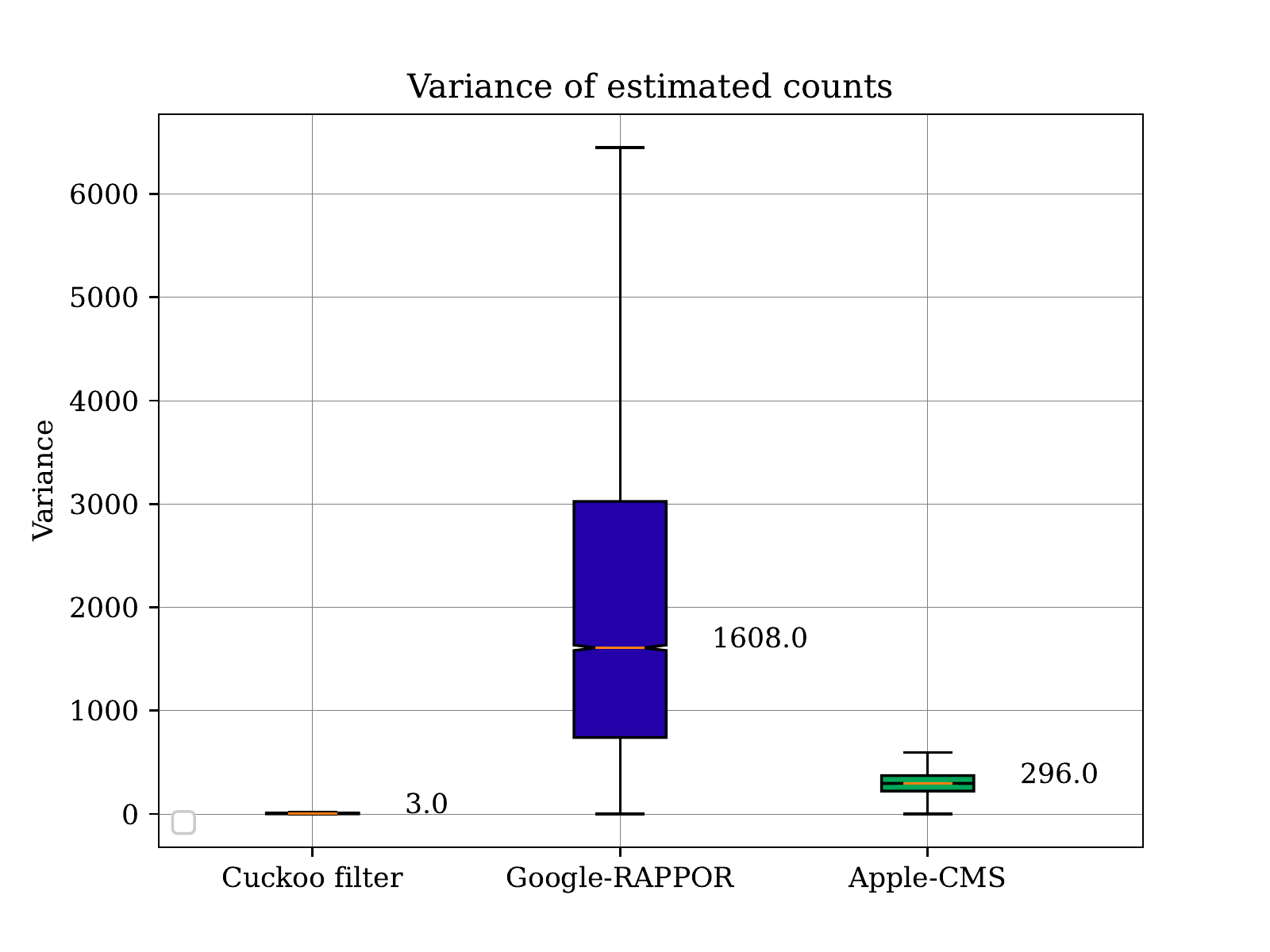}
 \includegraphics[width=0.38\textwidth]{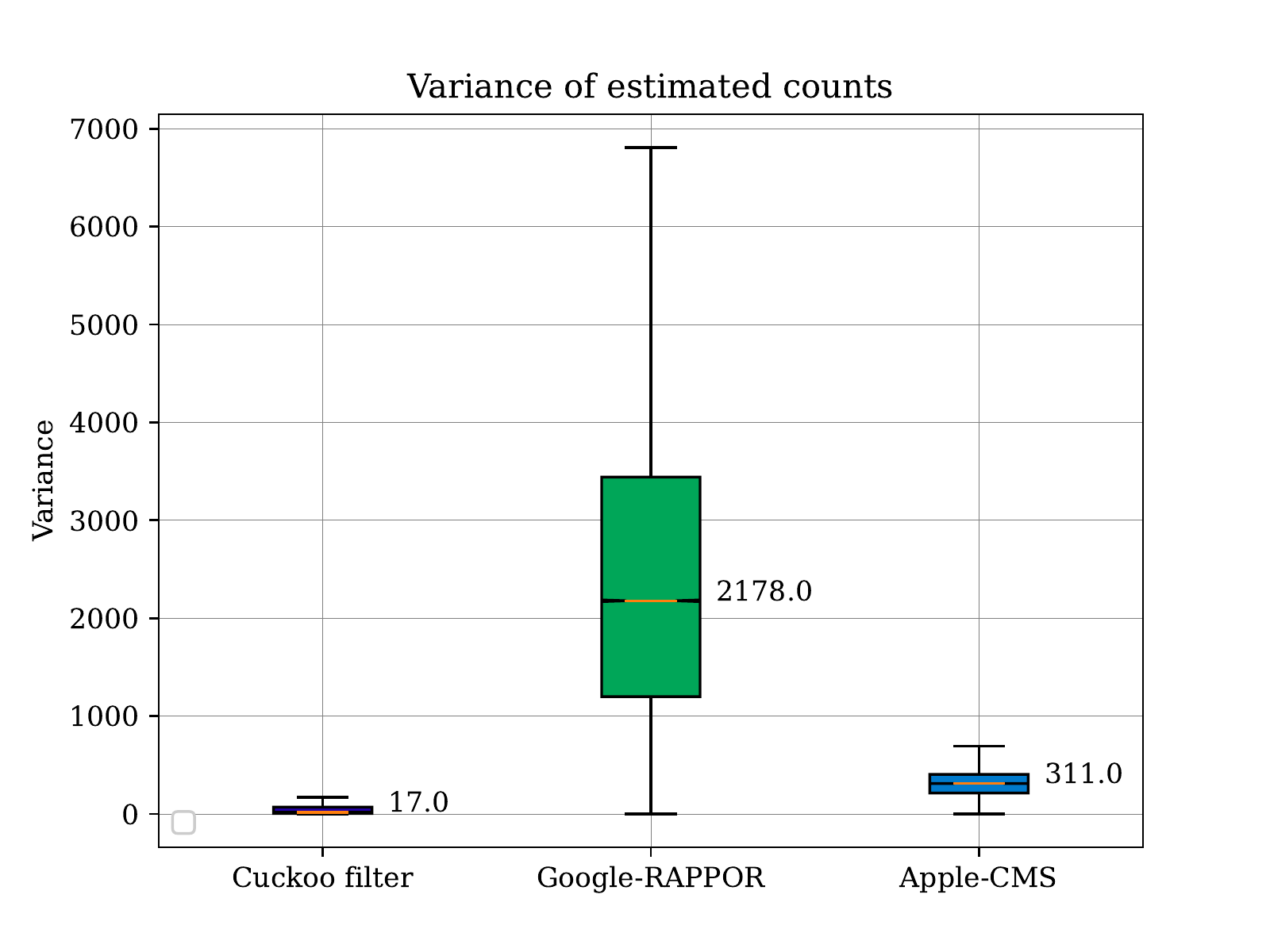} 
  \caption{\small{Comparison of effectiveness results of exact counting (left) and fuzzy counting (right) for  all approaches on the words (left) and synthetic words (right) datasets measured as the
  variance between actual and estimated counts of queried items.
  }
    }
\label{fig:comp_utility}
\end{figure*}

\begin{figure*}[ht!]
\centering
 \includegraphics[width=0.37\textwidth]{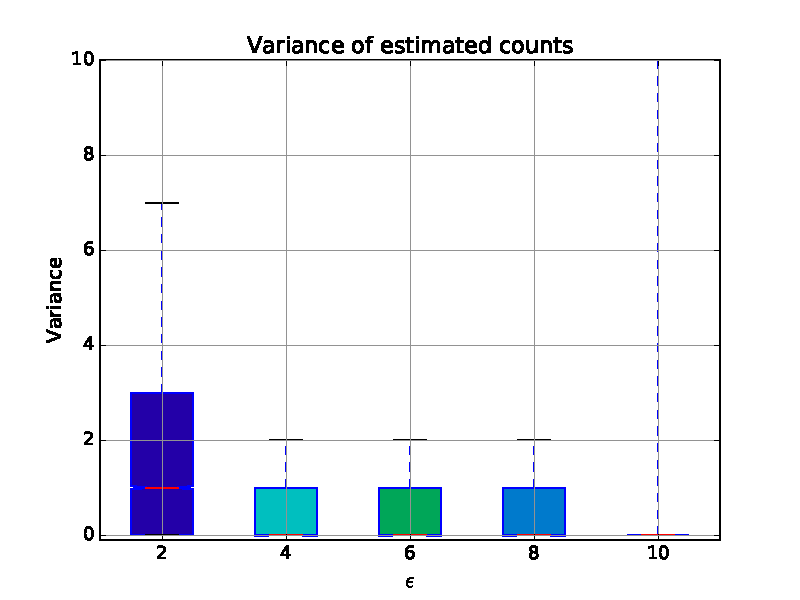}
 \includegraphics[width=0.38\textwidth]{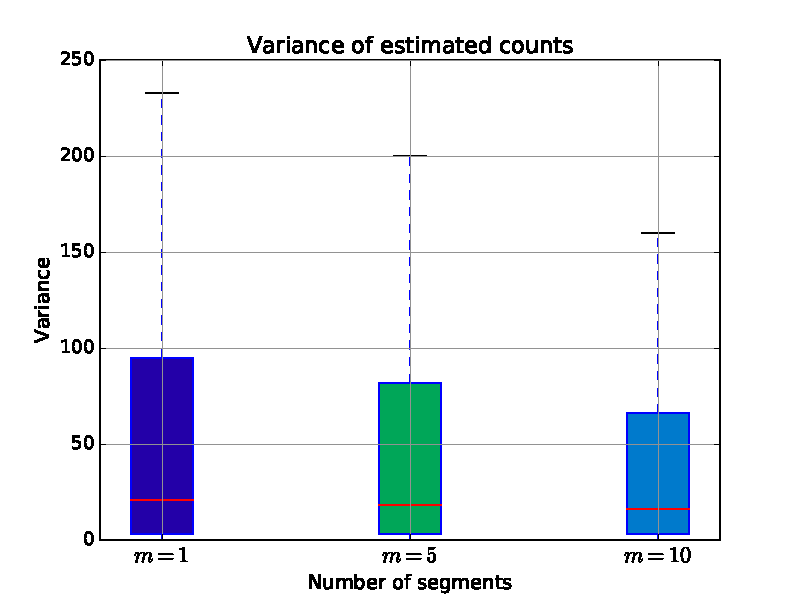} 
  \caption{\small{Ablation study: utility of count estimation of the proposed method measured as the median of the variance/error between actual and estimated counts of queried items in 10 iterations against different $\epsilon$ values on the words datasets (left) and against different number of segments $m$ on synthetic words datasets (right).
  }
    }
\label{fig:ablation_Study}
\end{figure*}

\subsection{Results and Discussion}
\label{subsec:results}
s
\noindent
\textbf{Efficiency results:}
We compare the insertion time and querying time of our method with the baselines for exact counting in Figures~\ref{fig:comp_efficiency}(left) and ~\ref{fig:comp_efficiency}(right), respectively. Please note that the insertion and querying time are shown in CDF plots, i.e. the total time required for inserting or querying a certain number of records (not a single record). The insertion time required for our method is slightly larger than the baseline approaches. The reason is that it needs to insert multiple fingerprints for every single item which requires several lookup operations in the Cuckoo filter even with parallel settings. This is negligible given the significant increment in utility of both exact and fuzzy counting with our method compared to the state-of-the-art methods, as will be discussed next. Moreover, the querying time is highly efficient and faster than the baselines (Fig.~\ref{fig:comp_efficiency} (right)). For real-time querying applications querying time is more important than insertion time, as the queries need to be instantly answered in real-time. Therefore, compared to RAPPOR and CMS, our method is highly applicable to real-time querying. 

We next present the querying time for fuzzy counting with our approach for different number of Bloom filter segments $m=1$, $m=2$, and $m=3$ in Fig.~\ref{fig:comp_efficiency_fuzzy} (left) on the synthetic words dataset where the queries contain data errors (e.g. typos) and variations. As expected, the querying time increases with larger $m$, as more segments need to be queried with larger $m$. Please note that parallel processing of insertion can help reducing the insertion time and hence the insertion time is not impacted with increasing $m$ (due to space limitation we do not show this results), but still all $m$ segments need to be queried from all cuckoo filters for a given query. The querying time for fuzzy counting is the same as for exact counting with RAPPOR and CMS. However, even with $m=5$, our cuckoo filter-based approach has considerably lower querying time than these state-of-the-art methods (i.e. faster querying time not only for exact counting, but also for fuzzy counting with a relatively larger number of segments). In other words, the querying time required for fuzzy counting by our method is still lower than the other two methods that only allow exact counting.

\noindent
\textbf{Effectiveness results:}
The result of the impact of continuous insertion of records on the utility or effectiveness of counting is illustrated in Figure~\ref{fig:comp_efficiency_fuzzy} (right). As can be seen, the false positive rate is not impacted with number of records inserted into the Cuckoo filter due to adaptively increasing the bucket size. Our method has significantly lower false positive rate (nearly $0.0$) compared to RAPPOR and Apple-CMS, and therefore provides higher quality of hash-mapping into the probabilistic data structure (Cuckoo filter). By bounding the false positive rate of adaptive Cuckoo filter to a very small value ($0.002$ in this experiment), the number of false positives that could occur due to hash-mapping in probabilistic data structures is negligible (therefore utility is only impacted by the differential privacy noise addition in our method).
Since RAPPOR uses cohorts to reduce the effect of false positives, it performs better than Apple-CMS until a certain number of records are inserted. 
These results reveal that our method can provide higher utility results with a significantly large margin even with large number of records inserted (in contrast to the baseline methods) at the cost of computational efficiency loss by a small margin. 

In Figure~\ref{fig:comp_utility} we evaluate the variance or absolute error of estimated and true counts of querying data with exact counting on words (left) and with fuzzy counting on synthetic words (right) datasets. Our results show that our method significantly outperforms the baselines by resulting in the lowest median of count estimation error in both cases, exact and fuzzy counting. As expected, RAPPOR estimates counts with large variance from true counts (due to the large false positive rate). Reducing the false positive rate by increasing the length of Bloom filters required extremely longer insertion and querying time for RAPPOR. The fuzzy counting results in Fig.~\ref{fig:comp_utility} (right) evaluate the fault tolerance/robustness of the methods to data errors and variations (using the synthetic words dataset). The significantly lower variance or error between estimated and true counts of our method on this dataset compared to the other two methods that do not support fuzzy counting validates the importance of fuzzy counting for improved accuracy of count estimation in data with errors and variations. 
In summary, our method achieves higher accuracy of count estimation compared to baselines by a significantly large margin for both exact and fuzzy counting. 

\noindent
\textbf{Ablation study:}
We performed an ablation study of our method with two important parameters of our method: a) privacy budget $\epsilon$ and b) number of Bloom filter segments for fuzzy counting $m$. The results are shown in Fig.~\ref{fig:ablation_Study}.
Fig.~\ref{fig:ablation_Study} (left) shows that the utility of count estimation by our method decreases with privacy budget ($\epsilon$). The trade-off between privacy and utility is reflected in these results.

Fig.~\ref{fig:ablation_Study} (right) shows the utility of fuzzy counting measured using the variance or absolute error between true and estimated counts with different $m$ values. As shown in the plot, with increasing $m$ the variance or error drops, i.e. the utility of fuzzy counting increases, as the cost of more querying time (as presented in Fig.~\ref{fig:comp_efficiency_fuzzy} (left). These results validate the trade-off between effectiveness and efficiency for fuzzy matching.

\section{Related Work}
\label{sec:rw}

Several privacy preserving counting or aggregation methods have been proposed in the literature. They can be categorized into three: 1) cryptographic methods, 2) sampling and statistics-based methods, and 3) probabilistic methods.

\noindent
\textbf{Cryptographic}: 
Searching on encrypted data for counting operations has been focused by several works~\cite{Bon04,Son00,Vo19}. However, most of these rely on the computation of very expensive functions (e.g. bilinear pairings) for each element/item in a dataset, rendering them not practical. An efficient and somewhat homomorphic encryption method, \textbf{EPiC}, using the MapReduce framework for cloud counting was proposed in ~\cite{Vo19}. However, these methods are not applicable to fuzzy counting or stream data applications.

PRIO is a privacy preserving system for the collection of aggregate statistics using cryptographic techniques~\cite{Cor17}. While the utility of aggregation with such cryptographic techniques is high, they do not allow fuzzy matching due to high computational cost required and do not support stream data processing. Some works considered combining multi-party computation (MPC) with differential privacy. For example, ~\cite{Pet15} developed secure aggregation protocols to add Laplace or Gaussian noise to ensure differential privacy for the clients, and ~\cite{Bin17} studied the overhead
of adding Laplace noise to MPC. 

\noindent
\textbf{Sampling and statistics}: 
Another category of methods for privacy preserving releasing or publishing of stream data considers providing statistics (e.g. moving average) of local data with local and/or central differential privacy (LDP or CDP) guarantees~\cite{Per19,Wan20}.
A recent work on answering range queries under LDP has proposed two approaches based on hierarchical histograms and the Haar wavelet transform to approximately answer range queries~\cite{Kul18}. Relaxed LDP was studied in~\cite{Xia20} that proposes E-LDP where E defines heterogeneous privacy guarantees for different pairs of private data values for significant utility gains in answering linear and multi-dimensional range queries. A recent work used sampling-based frequency estimation with fairness constraints which provides some level of privacy protection with good utility when the number of clients is small~\cite{Yan21}. A main challenge with these methods is that ensuring true randomness is a difficult task, so the success of random sampling is dependent on the data.

\noindent
\textbf{Probabilistic}: This category of methods received a lot of attention in the recent literature due to its computational and storage efficiency and provable privacy guarantees. The two state-of-the-art methods based on probabilistic methods for approximate aggregation are: (1) RAPPOR proposed by Google for anonymously collecting statistics about end-user software~\cite{Erl14} and (2) count-mean sketch (CMS) proposed by Apple for identifying frequently used words or emojis by users without compromising their privacy~\cite{Dp17} (described in detail in Section~\ref{sec:preliminaries}). 

Another method proposed by Apple to overcome the communication cost of CMS is the Hadamard transformation of CMS (HCMS)~\cite{Dp17}. Instead of sending over a vector to the server for inserting or updating an element, HCMS transforms the vector using Hadamard transformation and sends a single bit that is sampled from the transformed vector. Despite the improvement in the communication cost, it is not applicable to streaming data as the Hadamard transformation of the updated sketch is required upon every single item's insertion/update, which is computationally expensive for continuous data insertion and on-the-fly querying of frequency of items. 

Probabilistic data structures and differential privacy have also been used for other private counting tasks, such as counting of distinct elements~\cite{Sta17}, heavy hitters~\cite{Mel16,Kara15}, and incidence counting~\cite{Ala17}, due to their highly efficient computation, space, and communication costs.
Cuckoo filters have been widely used for private set intersection (PSI) with the purpose of increasing or optimising the efficiency (similar to blocking or hashing in record lookup or linkage techniques~\cite{Chr12}). For example, a PSI technique was proposed that utilizes Cuckoo hashing for batching~\cite{Che17}. Similarly, an efficient circuit-based PSI was proposed using Cuckoo hashing~\cite{Pin18}. Cuckoo filter was used to reduce the amount of data to be exchanged by a cryptography-based PSI~\cite{Res18}. Cuckoo hashing has been used for reducing the asymptotic overhead of Oblivious Transfer (OT)-based PSI protocol~\cite{Pin14}. This approach was further improved using permutation-based hashing and 3-way Cuckoo hashing where 3 instead of 2 hash functions are used to generate a densely populated Cuckoo table in order to reduce the overall number of OTs~\cite{Pin15}.

\section{Conclusion}
\label{sec:conclusion}

In this paper we have addressed a novel problem of privacy preserving real-time counting in stream data, which is increasingly being required in many real applications, using a hybrid method of Cuckoo filter and Bloom filter probabilistic data structures. Privacy preserving counting has been studied in the literature with two notable solutions by Google (RAPPOR) and Apple (CMS). We have investigated the additional challenges of answering real-time count querying in stream data and proposed a method that overcomes the limitations of existing methods for real-time counting in stream data applications.

We have proposed a novel local Differential privacy algorithm with low utility loss guarantees, and provided formal proof of privacy and utility guarantees of our method. We have conducted an experimental study of our method using large real and synthetic datasets. The results show that compared to two state-of-the-art approaches by Google and Apple, our method achieves significantly higher utility of count estimation and lower querying time at the overhead of higher insertion time with similar privacy guarantees. 

In the future, we aim to study noise addition according to ``local sensitivity" of the bucket in the Cuckoo filter in order to add noise based on the set of items the bucket has seen so far which could improve the utility and efficiency for a given privacy budget. The proposed differential privacy algorithm requires that clients need to keep track of the items reported to the server in order to avoid reporting the same item multiple times (i.e. only unique items are reported to the server). In the future, we aim to address the local storage/memory efficiency aspect by using efficient probabilistic data structures, such as HyperLogLog.

Further, investigating other counting tasks in stream data, including counting of distinct elements, heavy hitters, incidence counting, and estimating cropped means, is also important in different applications. For example, count-distinct is required to count views on a video, unique tweets, or unique customers accessed an online service. 
More work is required towards the development of advanced techniques for privacy preserving calculation of other statistical functions, such as conditional querying and incidence counting, in stream data. Investigating pan-private algorithms for streaming data applications is important to guarantee that the intermediate states in dynamic data are also private, i.e. the state of the algorithm after processing each iteration's update and the final output are jointly $\epsilon$-differentially private~\cite{Dwo10}.

\appendices

\section{Differential Privacy Proof for adjacency items being different items}
\label{app:proof}

\begin{theorem}[Differential privacy of Algorithm 2]
\label{th:dp}
Algorithm 2 is $\epsilon$-local Differentially private.
\end{theorem}

\begin{proof}
We will show that for an arbitrary set $V'$, the ratio of probabilities of observing $V'$ as output of Algorithm 2, given adjacent sets ${\cal D} = \{x_1\}$ and ${\cal D'} = \{x_2\}$ as input is bounded by $e^{\epsilon}$. Without loss of generality let $V' =[b_1, b_2, \cdots, b_u]$ denote the $u$ Bloom filters output by Algorithm 2. Then we want to show that, 

\begin{footnotesize}
\begin{equation}
    r=\frac{Pr[Algorithm~2(x_1) = [b_1, b_2, \cdots, b_u]]}{Pr[Algorithm~2(x_2) = [b_1, b_2, \cdots, b_u]]} \leq e^{\epsilon},
\label{eq:1}
\end{equation}
\end{footnotesize}

Note when ${\cal D} = \{ x_1\}$ is the input, at most one Bloom filter in $V'$ is due to item $x_1$, and that Bloom filter can be either from $x_1$'s true bucket (say $i_1$) or some other bucket $i_{\ell}$.  This is because the $generate\_similar\_bf(x_1)$ function can output a Bloom filter from either $x_1$'s true bucket or some other bucket. Thus, let's denote by $b_{i_1}$ and $b_{i_{\ell}}$ the respective Bloom filters that are added to $V'$ in those mutually exclusive cases. All other Bloom filters in $V'$ are the result of the flipping of $-1$ to $+1$ due to the randomized response (with parameter $p_f$) and the subsequent choosing of a random Bloom filter via the $randomly\_choose\_from()$ function.  Since the probability of these 'other' Bloom filters being in $V'$ is same whether the incoming item is $x_1$ or $x_2$, we only need to analyze the probability that input ${\cal D'} = \{ x_2\}$ results in the same Bloom filter $b_{i_1}$ (or $b_{i_{\ell}}$ as the case may be). 

For this analysis, we need to consider three buckets:  $i_1$ - the bucket of item $x_1$'s Bloom filter, $i_2$ - the bucket of item $x_2$'s Bloom filter and $i_{\ell}$ - the bucket of the 'similar' Bloom filter that is output by $generate\_similar\_bf(x_1)$.

We will now upper bound the ratio $r$ for the following three cases: 

Case 1: $b_{i_1}$ is in $V'$ due to item $x_1$.
In this case, when the incoming item is $x_1$ the bit at position $i_1$ remains $+1$ and the bits at $i_2$ and $i_{\ell}$ remain $-1$ after the RR mechanism and $generate\_similar\_bf(x_1)$ returns a Bloom filter from bucket $i_1$. While when the incoming item is $x_2$, the bit at position $i_1$ is flipped from $-1$ to $+1$, the bit at position $i_2$ is flipped from $+1$ to $-1$ and the bit at $i_{\ell}$ remains at $-1$ after the RR mechanism. Thus for this case the ratio $r$ is $r=\frac{(1-p_f).\frac{1}{2^{\ell(1-s_t)}}}{p_f.\frac{1}{t}} . \frac{(1-p_f)}{p_f} . \frac{(1-p_f)}{(1-p_f)}  $

Case 2: $b_{i_{\ell}}$ is in $V'$ due to item $x_1$.
In this case, when the incoming item is $x_1$ the bit at position $i_1$ remains $+1$ and the bits at $i_2$ and $i_{\ell}$ remain $-1$ after the RR mechanism and $generate\_similar\_bf(x_1)$ returns a Bloom filter from bucket $i_{\ell}$. While when the incoming item is $x_2$, the bit at position $i_1$ remains at  $-1$, the bit at position $i_2$ is flipped from $+1$ to $-1$ and the bit at $i_{\ell}$ is flipped from $-1$ to $+1$ after the RR mechanism.  Thus for this case the ratio $r$ is
$ r=\frac{(1-p_f).\frac{1}{2^{\ell(1-s_t)}}}{(1-p_f)} . \frac{(1-p_f)}{p_f} . \frac{(1-p_f)}{p_f.\frac{1}{t}}  $

Case 3: $\phi$ is in $V'$ due to item $x_1$.
In this case, when the incoming item is $x_1$ the bit at position $i_1$ is flipped to $-1$ and the bits at $i_2$ and $i_{\ell}$ remain $-1$ after the RR mechanism. While when the incoming item is $x_2$, the bit at position $i_1$ remains at  $-1$, the bit at position $i_2$ is flipped from $+1$ to $-1$ and the bit at $i_{\ell}$ remains at $-1$ after the RR mechanism.  Thus for this case the ratio $r$ is
$r = \frac{p_f}{(1-p_f)} . \frac{(1-p_f)}{p_f} . \frac{(1-p_f)}{(1-p_f)} $

Thus the ratio $r$ is upper bounded by $\frac{(1-p_f)^2.\frac{1}{2^{\ell(1-s_t)}}}{p_f^2.\frac{1}{t}} $. Thus $r \leq e^{\epsilon}$ implies $p_{flip} \geq \frac{1}{1+ \sqrt{s. e^{\epsilon}}}$

\end{proof}

\section*{Acknowledgment}
This research was funded by Macquarie University CyberSecurity Hub and strategic research funds from Macquarie University. Authors Dinusha Vatsalan and Raghav Bhaskar were affiliated with CSIRO Data61 at the initial submission of this manuscript.

%
%

\ifCLASSOPTIONcaptionsoff
  \newpage
\fi



\bibliographystyle{IEEEtran}

\bibliography{paper}
%

%

\begin{IEEEbiography}
[{\includegraphics[width=1in,height=1.25in,clip,keepaspectratio]{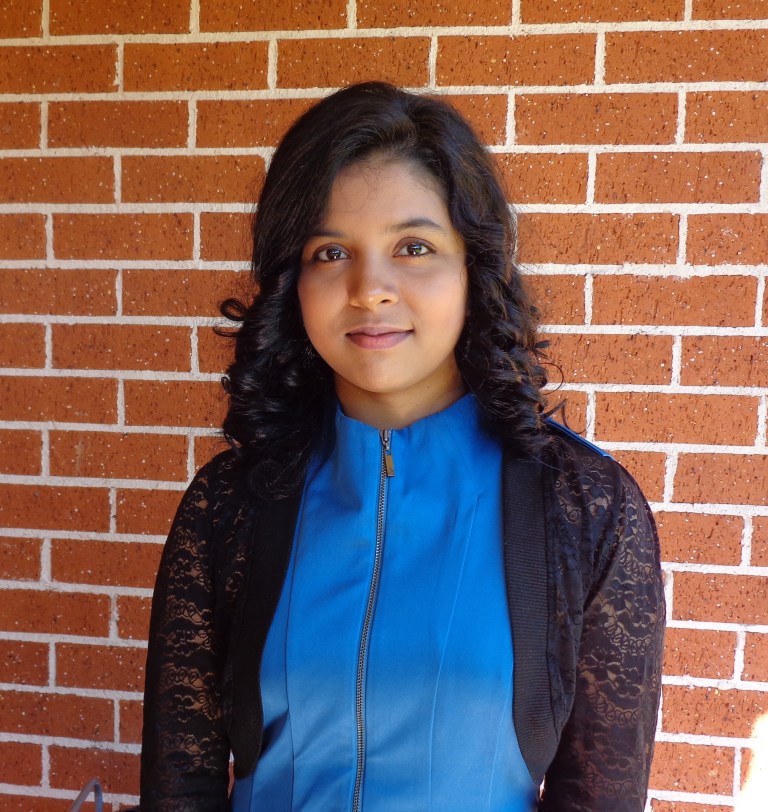}}]
{Dinusha Vatsalan} is a Senior Lecturer in Cyber Security at Macquarie University. Dinusha received her PhD in Computer Science from Australian National
University and BSc (Hons) from University of Colombo, Sri Lanka. She was
a Research Scientist at Data61, CSIRO.
Her research interests include privacy-preserving technologies for record linkage, data sharing, and analytics, privacy attacks and defences, and privacy risk quantification. She has authored over 60 scientific articles in these topics.
\end{IEEEbiography}

\begin{IEEEbiography}
[{\includegraphics[width=1in,height=1.25in,clip,keepaspectratio]{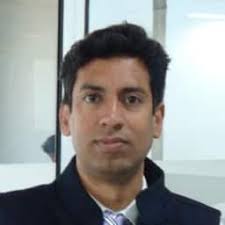}}]
{Raghav Bhaskar} is a Co-founder and CEO at AppsPicket, New Delhi, India with experience in the building blocks of cryptography (Digital signatures, Zero-Knowledge Proofs, Anonymous Credentials etc.) and privacy (Differential Privacy, Noiseless Privacy, Non-Interactive Zero Knowledge etc.) and interests in solving real life security and privacy challenges. He was a senior research scientist at Data61, CSIRO. 
\end{IEEEbiography}


\begin{IEEEbiography}
[{\includegraphics[width=1in,height=1.25in,clip,keepaspectratio]{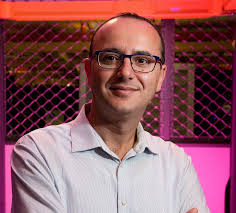}}]
{Mohamed Ali (Dali) Kaafar} is a Professor at the Faculty of Science and Engineering at Macquarie University and the Executive Director of the Optus-Macquarie University Cyber Security Hub. He is also the founder of the Information Security and Privacy group at CSIRO Data61. Prior to that, Dali was the research leader of the Data Privacy and Mobile systems group at NICTA and Senior principal researcher at INRIA, the French research institution of computer science and automation. 
He received his PhD from University of Nice Sophia Antipolis and INRIA in France where he pioneered research in the security of Internet Coordinate Systems. 

Prof. Kaafar is an associate editor of IEEE Transactions on Information Forensics \& Security and serves in the Editorial Board of the Journal on Privacy Enhancing Technologies. He published over 300 scientific peer-reviewed papers. 
He received several awards including the INRIA Excellence of research National Award, and the Andreas Pfitzman award from the Privacy Enhancing Technologies symposium in 2011. In 2019, he has also been awarded the prestigious and selective Chinese Academy of Sciences President's Professorial fellowship Award.
\end{IEEEbiography}




\end{document}